\documentclass[sigchi]{acmart}

\usepackage{graphics}
\usepackage{xspace}
\usepackage{color}
\usepackage{bm}
\usepackage{changes}
\usepackage{gensymb}
\usepackage{makecell}
\newcolumntype{C}[1]{>{\centering\arraybackslash}m{#1}}%
\colorlet{Changes@Color}{red}

\copyrightyear{2019}
\acmYear{2019}
\setcopyright{acmlicensed}
\acmConference[CHI 2019]{CHI Conference on Human Factors in Computing Systems Proceedings}{May 4--9, 2019}{Glasgow, Scotland UK}
\acmBooktitle{CHI Conference on Human Factors in Computing Systems Proceedings (CHI 2019), May 4--9, 2019, Glasgow, Scotland UK}
\acmPrice{15.00}
\acmDOI{10.1145/3290605.3300646}
\acmISBN{978-1-4503-5970-2/19/05}
\newcommand{\toolkitname}{\textit{OpenGaze}\xspace}

\def\plaintitle{Evaluation of Appearance-Based Methods and Implications for Gaze-Based Applications}

\def\plainkeywords{Appearance-Based Gaze Estimation; Model-Based Gaze Estimation; OpenGaze; Software Toolkit; Tobii EyeX}

\begin{document}
\title{\plaintitle}

\author{Xucong Zhang}
\affiliation{%
  \institution{Max Planck Institute for Informatics\\Saarland Informatics Campus}}
\email{xczhang@mpi-inf.mpg.de}

\author{Yusuke Sugano}
\affiliation{%
  \institution{Osaka University, Graduate School of Information Science and Technology}}
\email{sugano@ist.osaka-u.ac.jp}

\author{Andreas Bulling}
\affiliation{%
  \institution{University of Stuttgart, Institute for Visualisation and Interactive Systems}}
\email{andreas.bulling@vis.uni-stuttgart.de}
\email{}




\begin{abstract}
%!TEX root = 00_main.tex

Appearance-based gaze estimation methods that only require an off-the-shelf camera have significantly improved but they are still not yet widely used in the human-computer interaction (HCI) community.
This is partly because it remains unclear how they perform compared to model-based approaches as well as dominant, special-purpose eye tracking equipment.
To address this limitation, we evaluate the performance of state-of-the-art appearance-based gaze estimation for interaction scenarios with and without personal calibration, indoors and outdoors, for different sensing distances, as well as for users with and without glasses.
We discuss the obtained findings and their implications for the most important gaze-based applications, namely explicit eye input, attentive user interfaces, gaze-based user modelling, and passive eye monitoring.
To democratise the use of appearance-based gaze estimation and interaction in HCI, we finally present \textit{OpenGaze} (www.opengaze.org), the first software toolkit for appearance-based gaze estimation and interaction.
\end{abstract}

\maketitle

%!TEX root = 00_main.tex

\section{Introduction}

\begin{figure}[t]
\centering
  \includegraphics[width=\columnwidth]{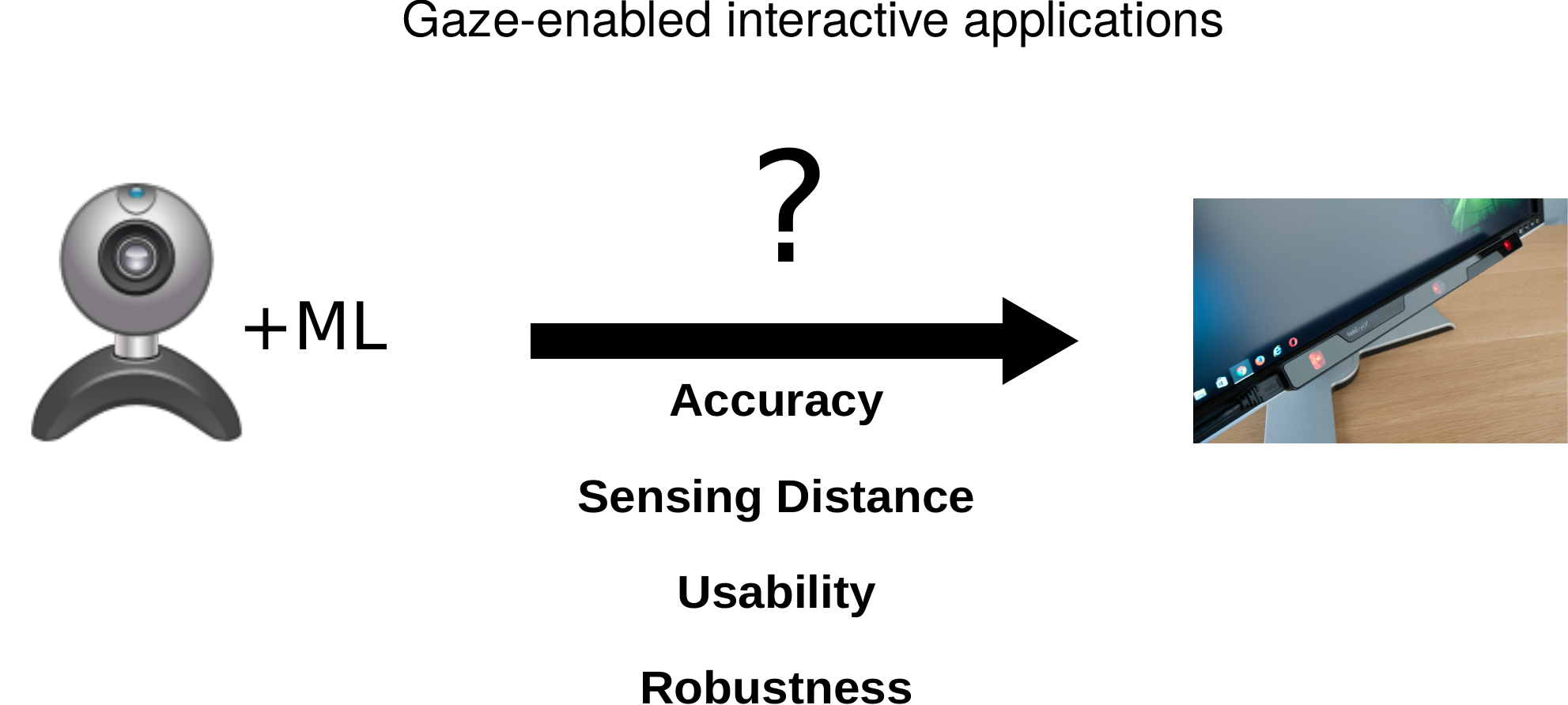}
  \caption{We study the gap between dominant eye tracking using special-purpose equipment  (right) and appearance-based gaze estimation using off-the-shelf cameras and machine learning (left) in terms of accuracy (gaze estimation accuracy), sensing distance, usability (personal calibration), and robustness (glasses and indoor/outdoor use).
  }
\end{figure}

Eye gaze has a long history as a modality in human-computer interaction (HCI), whether for attentive user interfaces~\cite{bulling2016pervasive}, gaze interaction~\cite{vidal13_ubicomp,zhang2017smartphone}, or eye-based user modelling~\cite{xu2016spatio,kosch2018your}.
A key requirement of gaze-based applications is special-purpose eye tracking equipment, either worn on the body (head-mounted) or placed in the environment (remote).
Despite the fact that the costs of hardware and software have decreased, particularly over the last couple of years, this requirement still represents a major hurdle for wider adoption of gaze in HCI research and practical applications.
Another hurdle is the need for expert knowledge on how to set up and operate these trackers to obtain accurate gaze estimates, i.e.\ how to calibrate them properly to each individual user.

With the goal to address these limitations, research in computer vision has focused on developing gaze estimation methods that are calibration-free and that only require off-the-shelf RGB cameras, such as those readily integrated in an ever-increasing number of personal devices or ambient displays~\cite{zhang2017mpiigaze,wood2014eyetab,huang2015tabletgaze}.
While model-based methods fit a geometric eye model to the eye image, appearance-based methods directly regress from eye images to gaze directions using machine learning~\cite{hansen2010eye}.
For a long time, these methods remained far inferior to special-purpose eye trackers, particularly in terms of gaze estimation error and robustness to head pose variations.
However, appearance-based gaze estimation methods have recently improved significantly~\cite{zhang17_cvprw,zhang2017mpiigaze} and promise a wide range of new applications, for example in attentive user interfaces~\cite{sugano2016aggregaze,zhang2017everyday}, mobile gaze interaction~\cite{khamis18_mobilehci} or social signal processing~\cite{mueller18_etra}.

Despite their potential, particularly given expected future improvements and availability of even larger-scale training data, appearance-based gaze estimation methods are still not yet widely used in HCI.
We believe this is partly because it currently remains unclear how they perform compared to dominant, special-purpose eye tracking equipment.
In the gaze estimation literature, evaluations have been often performed within the category that the proposed method belongs to.
We are not aware of a single, principled comparison of model- and appearance-based gaze estimation methods with a common stationary eye tracker.
Another likely reason is that using these methods remains challenging for user interface and interaction designers. 
While source code for some current methods is available~\cite{zhang17_cvprw,zhang2017mpiigaze,park18_etra}, the code has been written by computer vision experts for evaluation purposes.
The code is typically either not optimised for real-time use, doesn't implement all functionality required for interactive applications in a single pipeline, or cannot be easily extended or integrated into other software or user interface frameworks.

This work aims to provide a basis for developers from the HCI community to integrate the appearance-based gaze estimation method into interactive applications.
In order to achieve this goal, we make the following contributions:
First, we evaluate the accuracy of state-of-the-art appearance-based gaze estimation for interaction scenarios with and without personal calibration, indoors and outdoors, for different interaction distances, as well as for users with and without glasses.
We compare accuracy with a state-of-the-art model-based gaze estimation method~\cite{park18_etra} and, for the first time, with a commercial eye tracker.
Second, we discuss the obtained findings and their implications for the most important gaze-based applications~\cite{majaranta2014eye} ranging from explicit eye input, to attentive user interfaces and gaze-based user modelling, to passive eye monitoring.
Third, to democratise the use of appearance-based gaze estimation and interaction in HCI, we present \textit{OpenGaze}, the first software toolkit for appearance-based gaze estimation and interaction that is specifically developed for user interface designers.
The framework implements the full gaze estimation pipeline, is easily extensible and integratable, and is usable by non-experts.
%!TEX root = 00_main.tex

\section{Related Work}

We start with a general introduction of the various gaze-based interactions, followed by pertinent studies on gaze estimation methods.

\subsection{Gaze-based human-computer interaction}

Taking eye gaze from users as a command to computers is the most intuitive gaze-aware application.
The typical usage of eye gaze information is as a replacement of the mouse, such as typing words with the eye~\cite{mott2017improving}, indicating user attention~\cite{nguyen2016gaze}, and selecting items~\cite{zhang14_ubicomp}.
Researchers have investigated daily human-computer interactions using different eye movements, such as fixations~\cite{majaranta2009fast,higuch2016can,mott2017improving}, smooth pursuit~\cite{esteves2015orbits}, and eye gestures~\cite{mardanbegi2012eye}.

In addition, gaze information has also shown significant potential for user understanding.
Most intuitively, eye tracking techniques have been used to capture and infer user behaviours, such as eye contact~\cite{zhang2017everyday} and daily activities~\cite{bulling13_chi,steil2015discovery}.
Eye tracking data has been also used to recognise users' latent states, including interest and engagement \cite{li2017towards,lagun2014towards}, affective states \cite{muller2018detecting}, cognitive states~\cite{huang2016stressclick,matthews1991pupillary}, and attentive states~\cite{vertegaal2003attentive,faber2017automated}.
It has been pointed out that eye tracking data can even be associated with mental disorders, such as Alzheimer's disease~\cite{hutton1984eye}, Parkinson's disease~\cite{kuechenmeister1977eye}, and schizophrenia~\cite{holzman1974eye}.
Furthermore, eye tracking data holds rich personal information, including personality traits~\cite{hoppe2018eye}, gender~\cite{sammaknejad2017gender}, and user identity~\cite{cantoni2015gant}.

These gaze-base applications have been studied across different platforms, underlining the significance of gaze as an interaction modality, and a rich source of information on users as well as their mental and physical states, in both stationary and mobile settings.
The most prevalent examples include use on personal devices, such as desktops and laptops~\cite{zhang2018training,huang2016building}, tablets~\cite{zhang2018training,wood2014eyetab}, and mobile phones~\cite{huang2017screenglint}. 
More recently, new gaze-aware applications are emerging and eye tracking devices have been integrated in public displays~\cite{sugano2016aggregaze}, head-mounted VR devices~\cite{piumsomboon2017exploring}, and vehicles~\cite{palinko2010estimating}.
However, application scenarios have been still strongly influenced by the technical requirements of, mostly commercial, eye tracking devices.
The use of camera-based, in particular appearance-based, gaze estimation in interactive applications has not been fully explored due to the lack of a complete, extensible, and cross-platform software toolkit.

\subsection{Gaze estimation methods}

Gaze estimation methods can be categorised into feature-based, model-based, and appearance-based approaches~\cite{hansen2010eye}.
Feature-based methods use eye features for gaze direction regression, such as corneal reflections caused by reflections of an external light source on the cornea~\cite{zhu2005eye,zhu2006nonlinear}.
Feature-based methods are commonly used in commercial eye trackers, such as the entry-level eye tracker Tobii EyeX.

Model-based methods first detect visual features, such as pupil, eyeball centre and eye corners, and then fit a geometric 3D eyeball model to them to estimate gaze~\cite{chen20083d}.
While early model-based methods required high-resolution cameras and infrared light sources~\cite{ishikawa2004passive,yamazoe2008remote}, recent approaches only use input images from a single webcam~\cite{valenti2012combining,wood2014eyetab}.
More recent works leverage machine learning to improve the accuracy of eye feature detection, for example to train eye feature detectors with a large amount of synthetic data~\cite{baltrusaitis2018openface,park18_etra}.

Appearance-based methods also only require images obtained from an off-the-shelf camera, but directly learn a mapping from 2D input images to gaze directions using machine learning~\cite{tan2002appearance}.
Since there is no explicit eye feature detection step involved, this family of methods can typically handle input images with lower resolution and quality than model-based methods.
Recent works leveraged both large-scale training data and deep learning to significantly improve the gaze estimation accuracy in more challenging real-world settings~\cite{zhang2017mpiigaze,Shrivastava2016Learning,zhang17_cvprw}.
These advances have enabled a range of new applications, such as in eye contact detection~\cite{zhang2017everyday,mueller18_etra} or attention analysis on public displays~\cite{sugano2016aggregaze}.
Further new applications can be expected given the ever-increasing number of camera-equipped devices and displays, particularly mobile devices~\cite{khamis18_mobilehci}.

Mainly because these three families of gaze estimation methods have different requirements in terms of hardware and deployment setting, they have never been compared with each other in a principled way.
Consequently, interaction designers currently lack guidance on which methods they should choose for their particular applications.
As discussed above, this prevents the exploration of gaze interaction applications taking the full advantages of these different gaze estimation methods.
%!TEX root = 00_main.tex

\section{Dataset for Evaluation}
\label{sec:dataset}
In gaze estimation research in computer vision, the primary experiment of interest is typically a performance comparison between different gaze estimation methods~\cite{zhang2017mpiigaze,Shrivastava2016Learning}.
One likely reason for this is the lack of a suitable dataset that facilitates such a comparison.
We therefore collected a dataset specifically geared to study performance of the different methods with respect to core affordances important in gaze interaction research: specifically, the distance between user and camera, number of required calibration samples, use of the method indoors or outdoors, as well as whether the user wears glasses or not.

For data recording, we used a Logitech C910 webcam with a resolution of $1920 \times 1080$ pixels.
We chose Tobii EyeX as the representative for feature-based gaze estimation (commercial eye tracking) because it is affordable, has recently become popular and is used in a range of research (e.g.~\cite{kurauchi2016eyeswipe,schenk2017gazeeverywhere}).
Data collection was performed with 20 participants (10 female, aged between 21 and 45 years) whom we recruited through university mailing lists and notice boards.
Our participants were from six different countries, and four of them wore glasses during the recording.
During data collection, we labelled ground-truth gaze locations by showing the target stimuli on the screen as a circle shirking to a dot that the participants were instructed to look at.
The screen pose was measured using the mirror-based calibration~\cite{rodrigues2010camera} beforehand, and ground-truth gaze locations have been recorded in the 3D camera coordinate system.

\begin{figure}[t]
\centering
  \includegraphics[width=0.8\columnwidth]{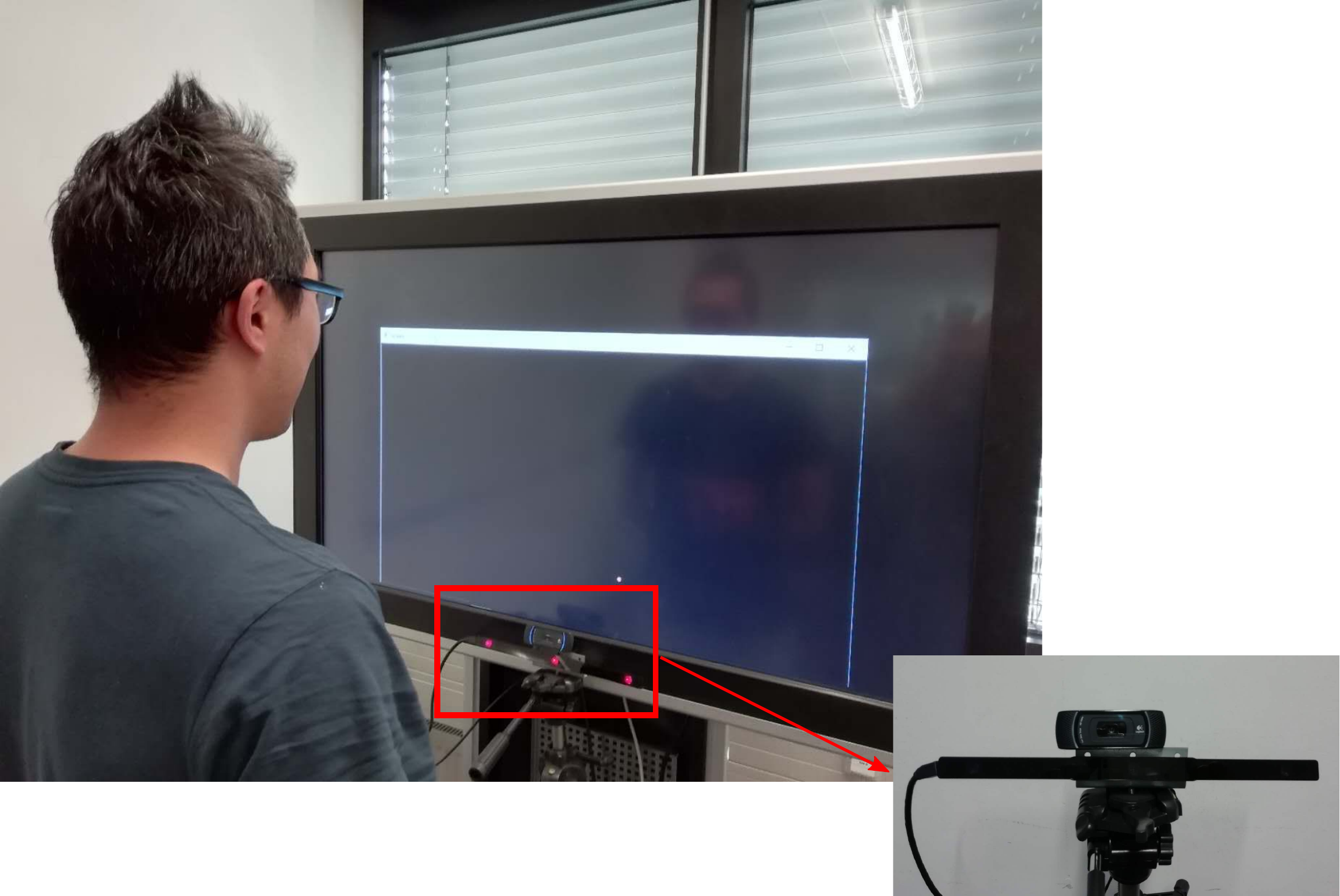}
  \caption{Our data collection setting. Participants stood at a pre-defined distance and were instructed to look at a dot on the screen. Data recording was performed using a Tobii EyeX eye tracker and a Logitech C910 webcam.}
  \label{fig:recording_setting}
\end{figure}

\paragraph{\textbf{Distances}}
Interaction distance is one of the most important factors to discuss the versatility of gaze estimation methods.
Distance can vary even for the same device, such as a mobile phone that is held at different distances, and even more so across different devices, such as a mobile phone or a public display.
Most commercial eye trackers, as represented by Tobii EyeX, are made to work optimally when the distance from the user is $50 - 90$ cm.
In contrast, webcam-based gaze estimation could output estimates for a large range of distances as long as the target faces are detected in the input image.
To evaluate the performance of gaze estimation methods at different distances, we collected the sample data with different distances between participants and cameras.
The recording setting is illustrated in~\autoref{fig:recording_setting}.
We showed the stimuli on a 55-inch public display, and mounted the webcam and Tobii EyeX below it.
We chose distances of 30, 50, 75, 110, 140 and 180 cm, where 50 and 75 cm fall inside the operational distance range of Tobii EyeX.
In order to make sure the ranges of view angles stay the same for different distances, the stimuli were displayed inside pre-defined regions corresponding to each distance.
They roughly correspond to 8.4, 12, 21, 32, 40 and 50 inches for 30, 50, 75, 110, 140 and 180 cm, respectively.

\paragraph{\textbf{Outdoor settings}}
\label{sec:outdoor_data}
Ideally gaze estimation methods should also yield robust performance independent of whether they will be used for interaction indoors or outdoors.
Therefore, we recorded two sessions with a laptop for both indoor and outdoor settings.
We mounted the Tobii EyeX below the laptop screen and a webcam above the screen due to the limited space.
We first instructed participants to stand or sit at around 50 cm from the cameras to collect the data, and repeated for both indoor and outdoor environments.
During recording for the outdoor setting, the participants were free to chose one of three locations outside our lab, and the recordings were done at different times of day.

\subsection{Procedure}\label{sec:procedure}

During recording, participants were asked to stand or sit at certain distances, and were instructed to look at a shrinking circle at random locations and click the mouse when the circle became a dot.
For each distance, we collected 80 samples:
60 samples were used for personal calibration and the rest for testing.
We continuously recorded video stream with a webcam, together with gaze estimates from the Tobii EyeX.
The time stamps were logged individually for mouse clicking events, video frames, and outputs of the Tobii EyeX.

Since the calibration procedure implemented in the Tobii SDK is black-box and could be different from our implementation, we collected samples both with and without personal calibration from the Tobii SDK for comparison.
Specifically, for distances of 50 and 75 cm, we first recorded 20 samples with the calibration profile from another person.
These samples were used as a test set for the Tobii EyeX without any personal calibration.
Then we performed the personal calibration provided by the Tobii SDK, which includes seven calibration points.
Finally, we recorded the 80 samples with Tobii EyeX together with the webcam.

\begin{figure*}[t]
\centering
  \includegraphics[width=0.9\textwidth]{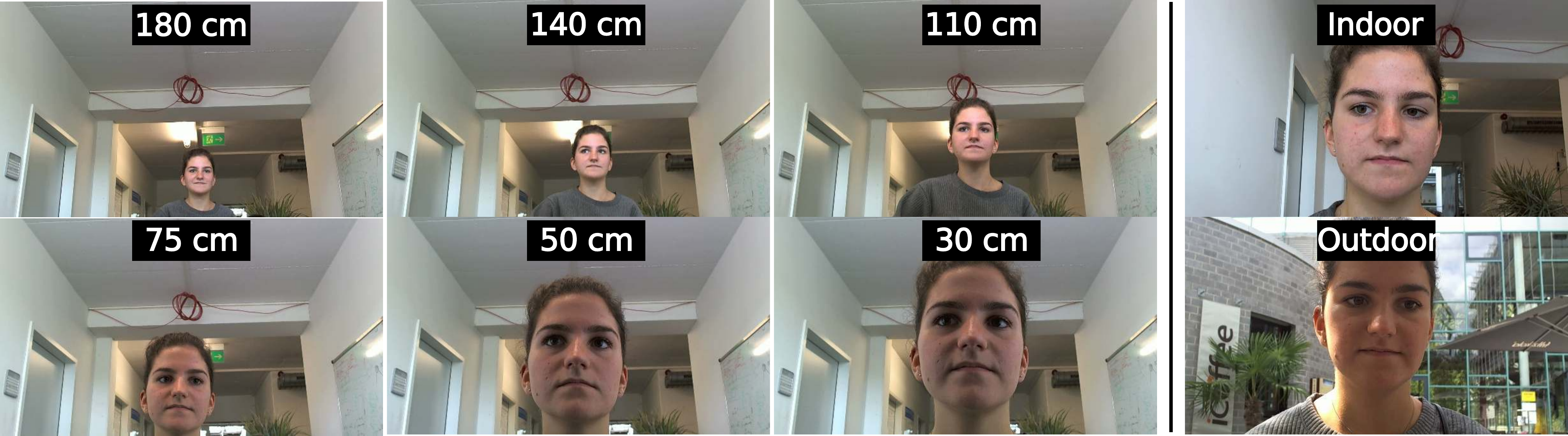}
  \caption{Data samples from one of our participants. Left: samples with recording distances marked at the top of the images. Right: samples under indoor and outdoor conditions, as marked at the top of images. As can be seen, our recorded data includes varying face sizes caused by different distances, and illumination conditions in indoor and outdoor settings.}
  \label{fig:data_samples}
\end{figure*}

\autoref{fig:data_samples} shows example images recorded from the webcam at different distances, as well as in indoor and outdoor settings.
The recording distances and conditions are marked on top of the samples.
As can be seen from these images, the distances lead to different face sizes, which can affect the input quality for gaze estimation methods.
The indoor and outdoor settings have very different illumination conditions, which directly affected the appearance of the faces.
Besides, the sunlight in the outdoor setting affects the active infrared light of Tobii EyeX, which resulted in as more invalid gaze data compared to the indoor setting.

%!TEX root = 00_main.tex

\section{Experiments}

The main goal of our experiments was to study the accuracy gap of state-of-the-art appearance-based gaze estimation (represented by MPIIFaceGaze~\cite{zhang17_cvprw}) with a model-based counterpart (represented by GazeML~\cite{park18_etra}) as well as a commercial eye tracker (Tobii EyeX).
The GazeML pupil detector was trained on large-scale synthetic eye images~\cite{wood2015rendering} with deep convolutional neural networks.
The method takes the vector from the estimated 3D eyeball centre to the detected pupil location as the estimated gaze direction.
We therefore deemed GazeML to represent the state of the art in model-based gaze estimation, because it reported (confirmed in our own comparisons) better accuracy than the gaze estimation method~\cite{wood2015rendering} implemented in the widely used OpenFace toolkit~\cite{baltrusaitis2018openface}.
We trained the MPIIFaceGaze method using two commonly-used gaze datasets with full-face images, MPIIFaceGaze dataset~\cite{zhang17_cvprw} and EYEDIAP dataset~\cite{mora2014eyediap}.
According to the training data distribution, this pre-trained model can handle head poses between $[-25\degree, 25\degree]$ horizontally and $[-10\degree, 25\degree]$ vertically under challenging real-world illuminations.

Eye tracking accuracy is often measured in terms of 2D gaze estimation error on the screen calculated as the differences between ground-truth and estimated gaze direction.
However, since 2D on-screen error measurement also depends on the distance between the screen and user, it cannot be used to compare accuracy on our data with varying distances.
Instead, we measured the 3D gaze estimation error in degrees, i.e.\ the difference between the estimated and the ground-truth 3D gaze vectors.
2D gaze points on the screen can be converted to 3D vectors in the camera coordinate system by using the screen-camera relationship.
The on-screen gaze location represents the point end of the gaze vector, while the face centre serves as the starting point~\cite{zhang17_cvprw}.
Given that MPIIFaceGaze can output gaze vectors in the camera coordinate system, those can be directly compared with the ground truth vectors.
GazeML outputs two gaze vectors, one for each eye.
We first projected them to the screen plane to obtain two intersecting points, and then took the middle point of the two intersection points as the point end of the gaze vector.
Tobii EyeX outputs 2D on-screen locations that were used as the point end of the gaze vector.

\subsection{Distances between user and camera}

\begin{figure}[t]
\centering
  \includegraphics[width=0.9\columnwidth]{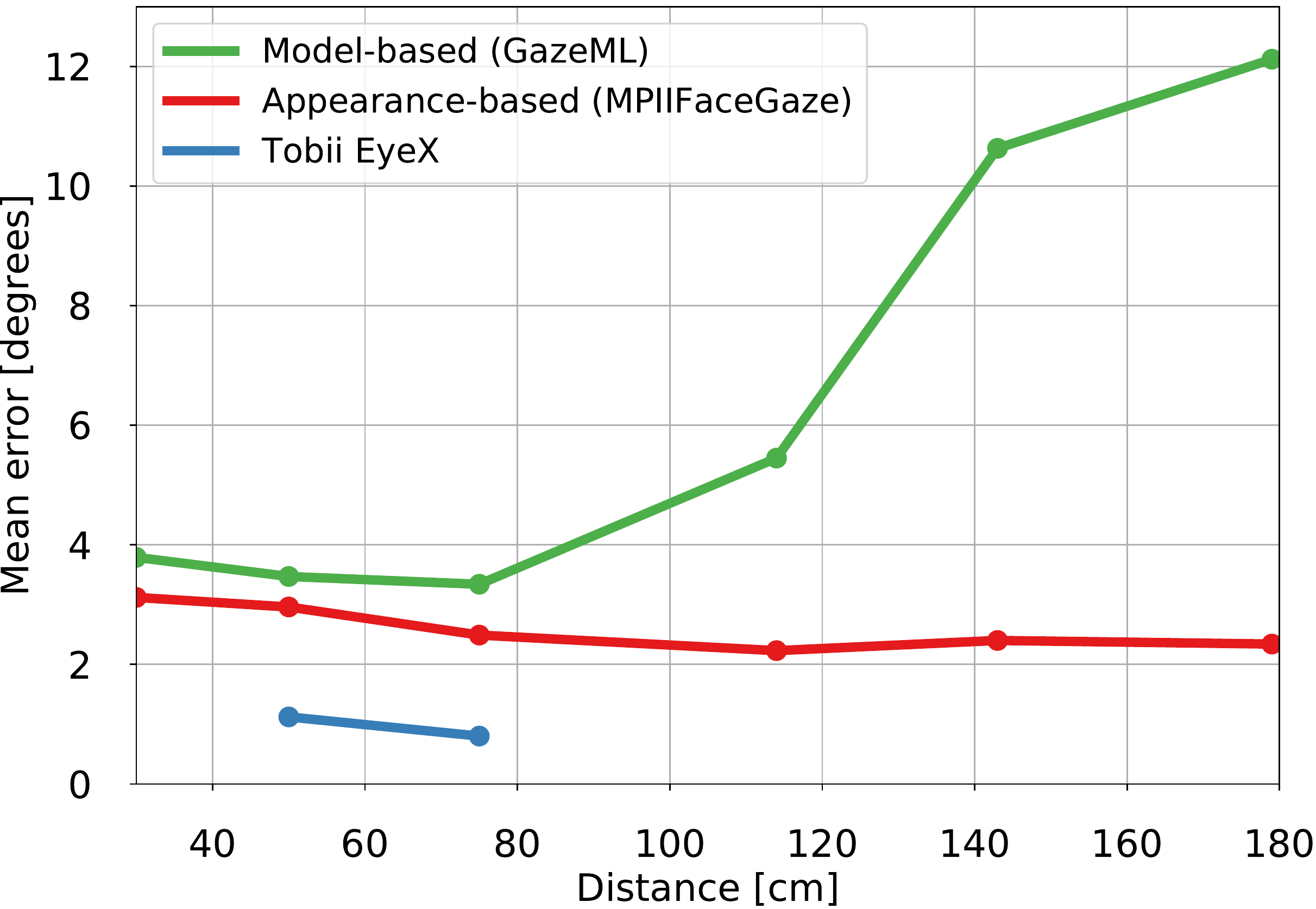}
  \caption{Gaze estimation errors of different methods in degrees across distances between the user and camera. Dots are results averaged across all 20 participants for each distances, and we linked them by lines.}
  \label{fig:results_distances}
\end{figure}

We first evaluated accuracy of the different methods across different distances between user and camera.
With our recorded data, we used all of the 60 samples to perform personal calibration for all methods, and tested them on the remaining 20 samples.
To show the full ability of Tobii EyeX, we first used its own personal calibration provided by Tobii SDK which requires seven calibration points.
We then applied the personal calibration with an additional 53 samples. 
We conducted this calibration procedure at each distance.
The errors were averaged across all participants.

The results are summarised in~\autoref{fig:results_distances}.
There are statistically significant differences between the three methods (t-test, $p<0.01$).
While Tobii EyeX performed the best with 1.2 degree gaze estimation error for distance 50 cm and 0.8 degrees for distance 75 cm, the tracking range of Tobii EyeX is severely limited compared to the other methods.
The appearance-based method (MPIIFaceGaze) achieved the second-best result as mean gaze estimation error from 2.3 degrees to 3.1 degrees, and this accuracy is robust across the full distance range from 30 cm to 180 cm with only minor variation.
The model-based method (GazeML) achieved the worst accuracy with mean gaze estimation errors ranging from 3.8 degrees to 12.1 degrees.
In contrast to MPIIFaceGaze, GazeML's accuracy was also sensitive to the distance: the larger the distance, the worse its accuracy.
This is most likely caused by the fact that accurate pupil detection and eyeball centre estimation, on which these types of methods crucially rely, become increasingly difficult with larger distances.

In summary, this evaluation shows that while there is still a accuracy gap of around two degrees between the appearance-based method and Tobii EyeX, the former has a much larger operational range.
This finding underlines the practical usefulness of appearance-based gaze estimation, in particular for interactive applications where the ability to track robustly across a large interaction space is important and where gaze estimation error can be compensated for, e.g.\ on interactive public displays using pursuits~\cite{vidal13_ubicomp}.

\subsection{Number of calibration samples}

\begin{figure}[t]
\centering
  \includegraphics[width=0.9\columnwidth]{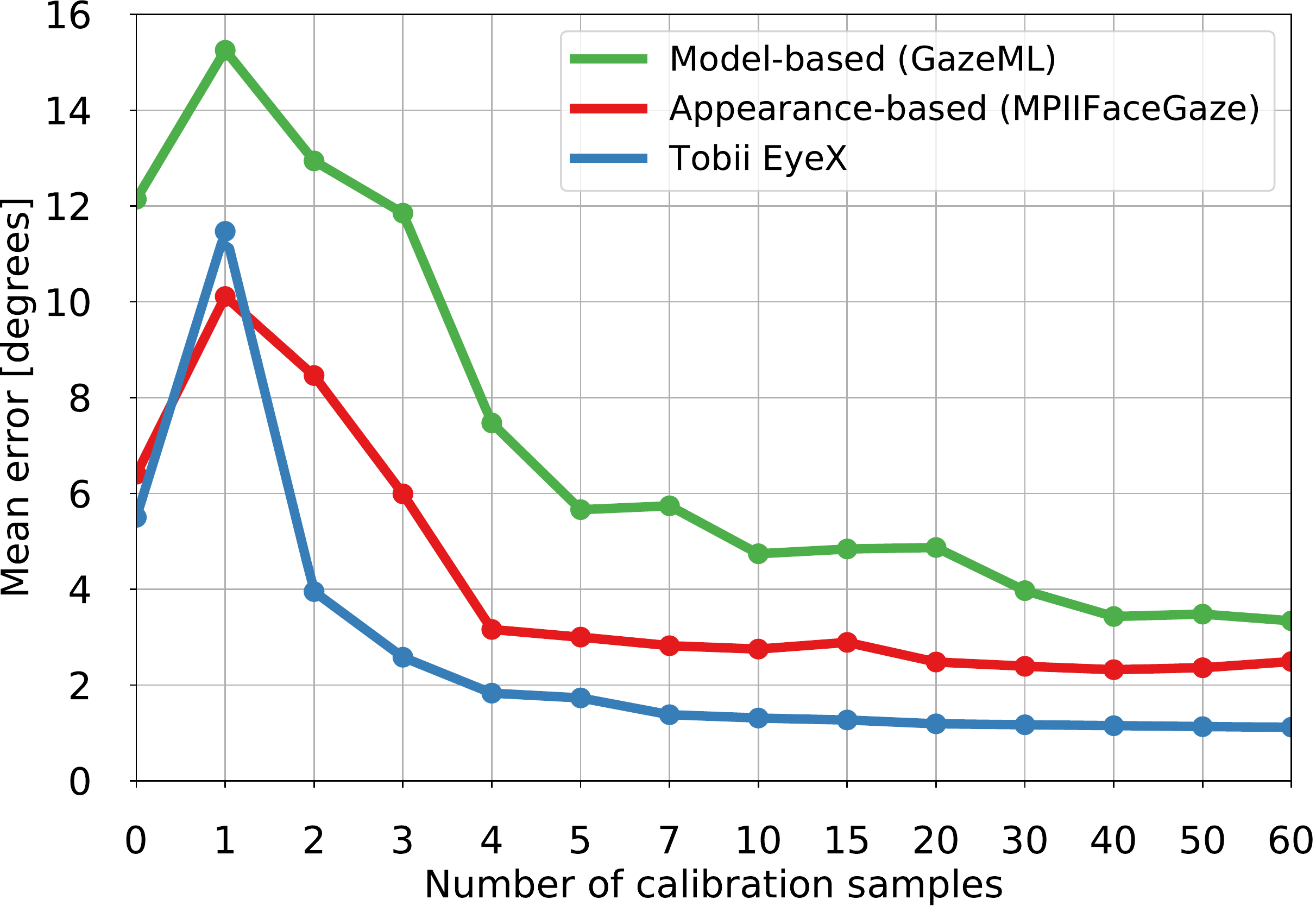}
  \caption{Gaze estimation errors of different methods in degrees across numbers of personal calibration samples. Dots are results averaged across all 20 participants, and we linked them with lines.}
  \label{fig:results_samples}
\end{figure}

The number of required calibration samples is an important factor for usability.
Calibration with a large number of samples can be time-consuming and prohibitive for certain applications where spontaneous interaction is crucial, e.g. on gaze-enabled public displays~\cite{zhang13_chi}.
Therefore, we evaluated the influence of the number of calibration samples on gaze estimation accuracy.
We analysed accuracy while varying the number of samples used for calibration, from zero, to one, two, three, four, five, seven, 10, 15, 20, 30, 40, 50 and 60.
For the calibration-free case (zero calibration samples) we directly used the raw gaze estimates as calculated by GazeML and MPIIFaceGaze, while Tobii EyeX was uncalibrated.
We opted for a distance of 75 cm because this is exactly within the optimal tracking range for Tobii EyeX.

\autoref{fig:results_samples} summarises the results and reveals a number of interesting insights.
As expected, the calibration-free setting achieves a large gaze estimation error, between 5.5 degrees of visual angle (for Tobii EyeX) and 12.1 degrees (for model-based GazeML).
The appearance-based method (MPIIFaceGaze) achieved 6.4 degrees.
These differences were statistically significant (t-test, $p<0.01$).
However, accuracy gets even worse for one-calibration samples, where the third-order polynomial mapping function is underdetermined.
With an increasing number of calibration samples, gaze estimation error decreases for all methods, down to 1.1 degrees for Tobii EyeX, and 2.5 degrees for MPIIFaceGaze.
These results show that current appearance-based methods (MPIIFaceGaze) can achieve accuracy competitive with Tobii EyeX, even with only four calibration samples.
This is exciting given that current appearance-based methods are competitive in terms of accuracy and usability and, hence, seem feasible for a range of everyday gaze interfaces, such as on camera-equipped personal devices.

\subsection{Indoor and outdoor settings}

\begin{figure}[t]
\centering
  \includegraphics[width=0.9\columnwidth]{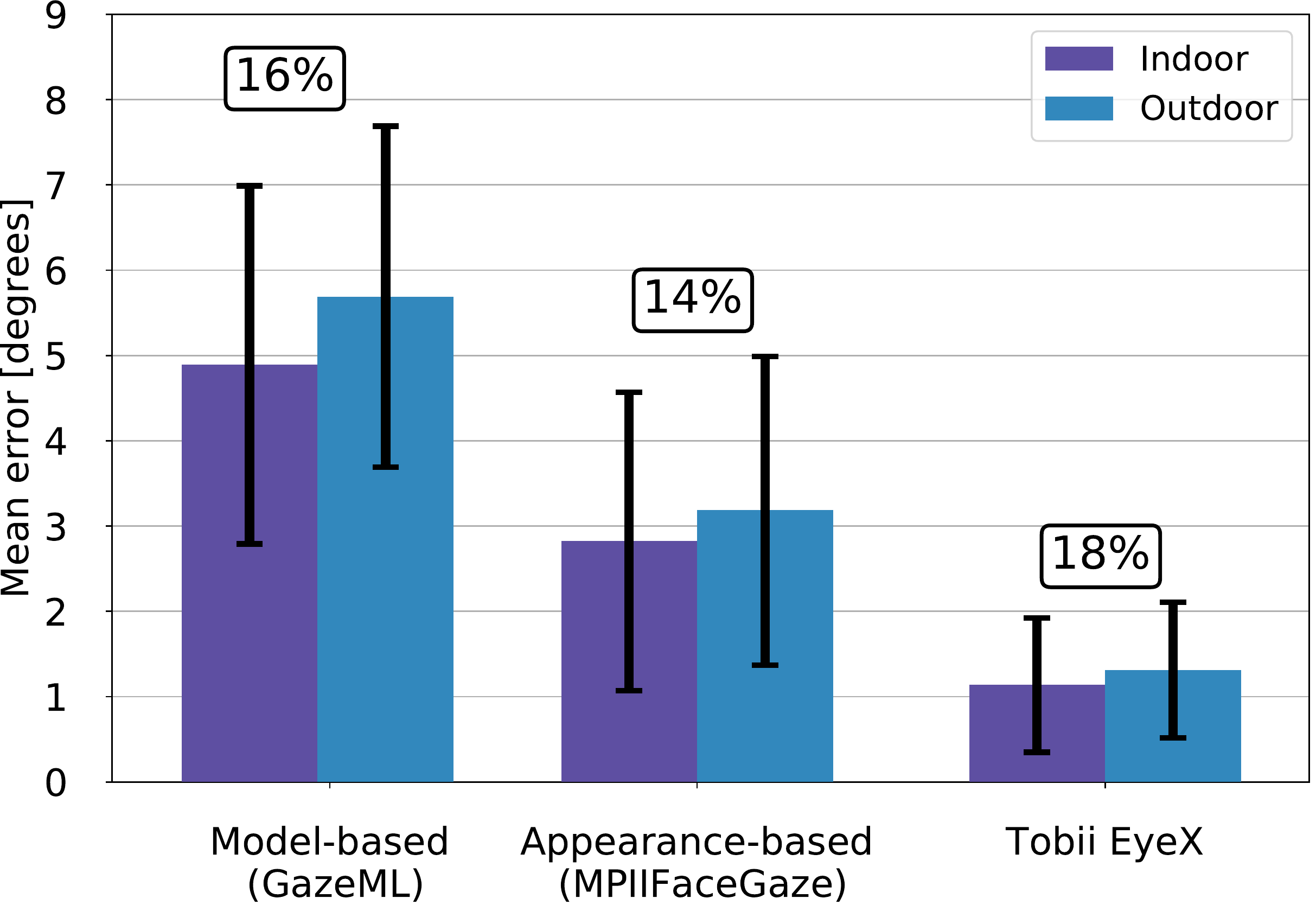}
  \caption{Gaze estimation errors for indoor and outdoor settings. Bars show mean error across all participants; error bars indicate standard deviations. The numbers above the bars indicate accuracy differences from indoor to outdoor setting in percent.}
  \label{fig:results_indoor_outdoor}
\end{figure}

We then evaluated the impact on accuracy of different illumination conditions -- a common problem when moving between indoor and outdoor interaction settings.
For this evaluation, we used the data collected for indoor and outdoor settings described in~\autoref{sec:dataset}.
We again used all 60 samples for personal calibration and evaluated the different methods for both settings. 
The results of this evaluation are summarised in~\autoref{fig:results_indoor_outdoor} where bars show the gaze estimation error in degrees across all 20 participants, and error bars show standard deviations.
The figure also shows the relative accuracy differences between indoors to outdoors in percent.

As can be seen from the figure, the best accuracy was again achieved by Tobii EyeX (indoors: 1.1 degrees, outdoors: 1.3 degrees), followed by the appearance-based method (MPIIFaceGaze) (indoors: 2.8 degrees, outdoors: 3.2 degrees), and model-based method (GazeML) (indoors: 4.9 degrees, outdoors: 5.7 degrees).
Although the accuracy differences between indoors and outdoors are not statistically significant (t-test, $p>0.05$), better accuracy tends to be achieved for the indoor environment, likely as a result of changing illumination conditions.

\subsection{With and without glasses}

\begin{figure}[t]
\centering
  \includegraphics[width=0.9\columnwidth]{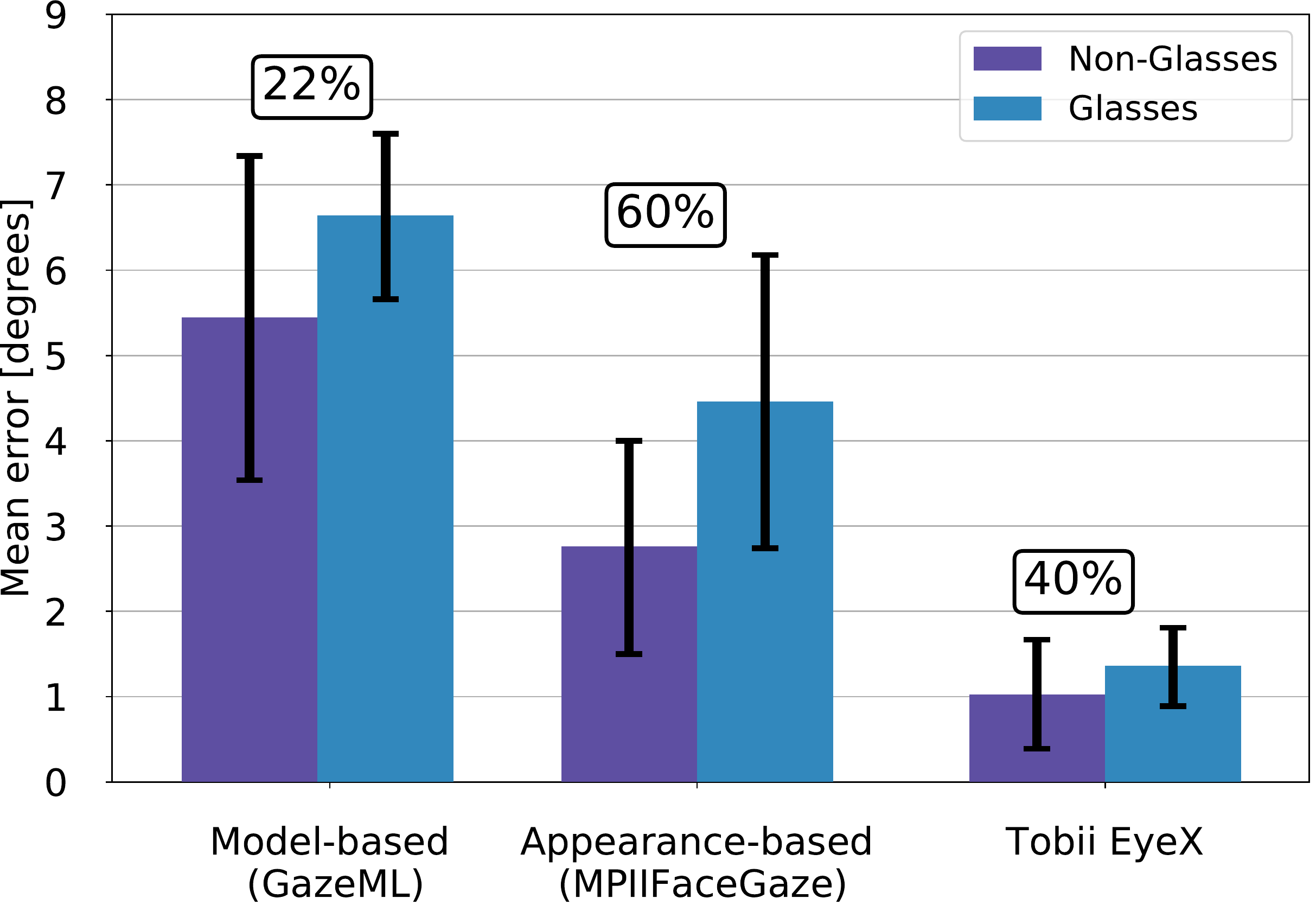}
  \caption{Gaze estimation errors for participants wearing glasses or not. Bars show mean error across all participants; error bars indicate standard deviations. The numbers above the bars indicate accuracy differences between not wearing glasses and wearing glasses in percent.}
  \label{fig:results_glasses}
\end{figure}

Glasses can have a significant effect on gaze estimation error due to strong reflections and distortions they may cause.
In our dataset, four participants wore glasses while the rest did not.
To evaluate the impact of glasses, we analysed the error after personal calibration with 60 samples, distances range ranging 50 and 75 cm, and in both indoor and outdoor settings.
The results of this evaluation are shown in~\autoref{fig:results_glasses}, where bars show the gaze estimation error in degrees across participants when wearing or not wearing glasses, and the error bars show standard deviations.
These differences were statistically significant (t-test, $p<0.01$).
The figure also shows the relative accuracy differences between wearing glasses and not wearing glasses in percent.

As we can see from the figure, glasses have a stronger effect on gaze estimation accuracy than illumination conditions (see \autoref{fig:results_indoor_outdoor}).
The gaze estimation errors were 5.4 (without) and 6.3 (with) degrees for the model-based method (GazeML), 2.8 and 4.5 degrees for the appearance-based method (MPIIFaceGaze), and 1.0 and 1.4 degrees for Tobii EyeX.
The estimation results differences between with and without glasses are larger for the appearance-based method (MPIIFaceGaze) than for Tobii EyeX.
Similarly, as for the previous evaluation on indoor and outdoor settings, the likely reason for this is the training data that, in this case, does not contain a sufficient number of images of people wearing glasses.
As a result, the appearance-based method cannot handle these cases as well.
Another reason could be that Tobii EyeX uses infrared light, which filters out most reflections on the glasses.

%!TEX root = 00_main.tex

\section{Implications for gaze applications}
Gaze estimation devices, especially the dominant commercial eye trackers, have facilitated the development of gaze-based interactive systems in past years.
In this section, we discuss the implications of our findings for the most important gaze-based applications as well as the potential of appearance-based methods for application scenarios that only require a single off-the-shelf camera.

\subsection{Gaze applications}
Gaze-based interactive applications can be divided into four groups: \textit{explicit eye input}, \textit{attentive user interfaces}, \textit{gaze-based user modelling}, and \textit{passive eye monitoring}~\cite{majaranta2014eye}.
The \textit{explicit eye input} applications take the gaze input to command and control the computer.
\textit{Attentive user interfaces} do not expect explicit commands from the user,  while using the natural eye movements subtly in the background.
\textit{Gaze-based user modelling} uses gaze information to understand user behaviour, cognitive processes, and intentions; this usually utilise short-time-period data.
\textit{Passive eye monitoring} stands for off-line analysis with long-term gaze data for diagnostic applications.

\begin{figure}[t]
\centering
  \includegraphics[width=\columnwidth]{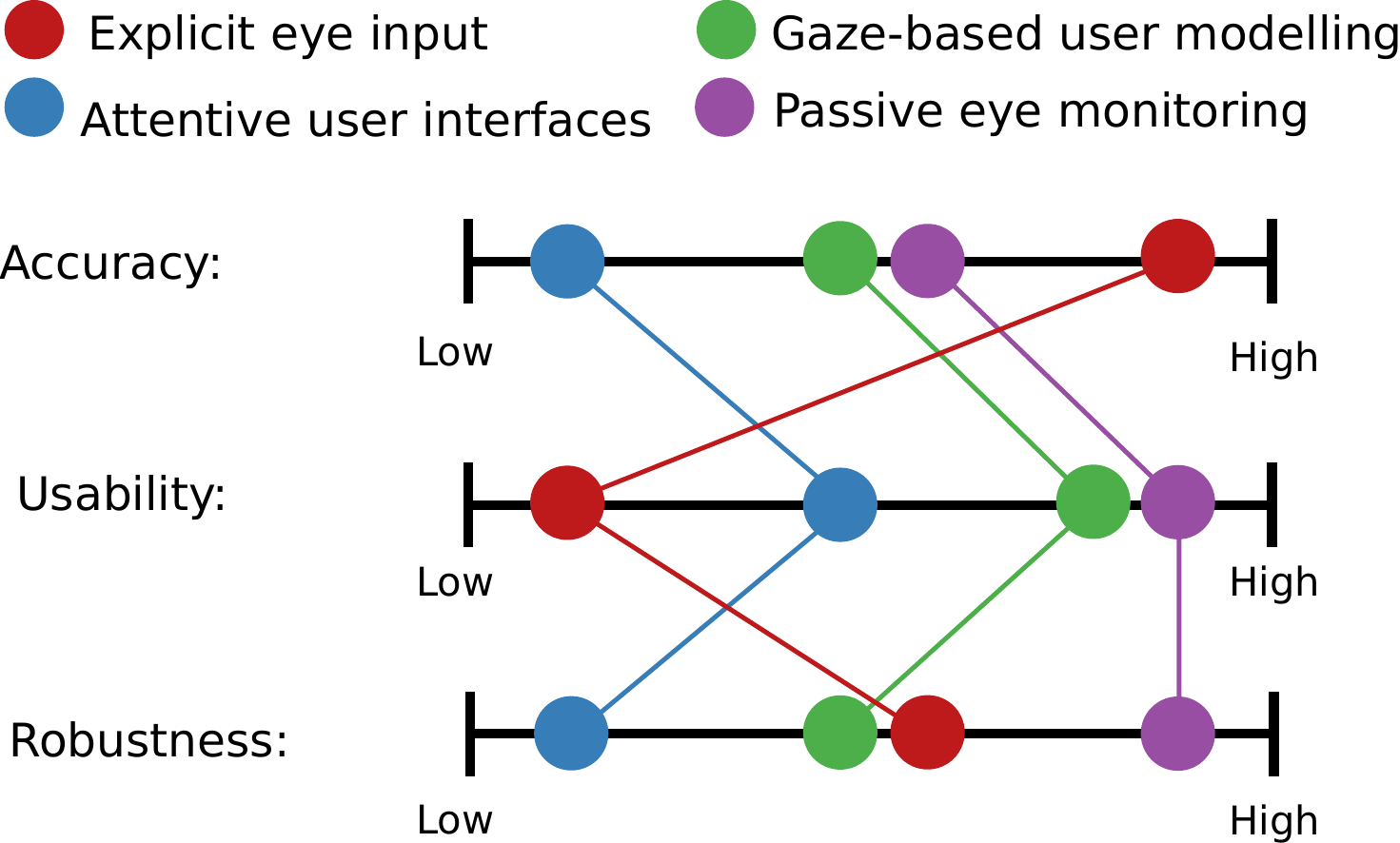}
  \caption{Relationship of different gaze-based applications and affordances. We show three typical gaze-based applications and the different levels of each affordance.}
  \label{fig:properties}
\end{figure}

While all of the above categories take gaze information from users; they have different requirements in terms of properties on the gaze estimation methods.
In this section, we summarise the requirements of applications regardless of the technical limitations of existing eye tracking methods.
In this way, we clarify how the application scenarios can be extended by using appearance-based gaze estimation beyond the limitation of commonly-used commercial eye trackers.
We show the relationships of different gaze-based applications and affordances in~\autoref{fig:properties}, and explain them in the following.

\paragraph{\textbf{Accuracy}}
Applications that rely on \textit{explicit eye input} usually require high-accuracy gaze estimates, such as for eye typing~\cite{kurauchi2016eyeswipe,mott2017improving}, authentication~\cite{khamis16_chi}, or system control~\cite{nguyen2016gaze}.
However, the allowed gaze estimation error depends on the sizes of the gaze targets.
These gaze targets could be fine-level details on the screen~\cite{d2018eye}, closely connected~\cite{zhang2017look} or separate content on the screen~\cite{d2016gazed}, large physical objects~\cite{andrist2017looking}, or rough gaze direction ~\cite{otsuki2017thirdeye,zhang2017smartphone}.
\textit{Gaze-based user modelling} and \textit{passive eye monitoring} require gaze estimation to detect gaze patterns instead of individual points.
These gaze patterns could be large regions on the screen, such as during saliency prediction~\cite{xu2016spatio}; relative eye movements for inferring everyday activities~\cite{steil2015discovery,sattar15_cvpr}, cognitive load and processes~\cite{tessendorf11_pervasive,bulling14_pcm}, or mobile interaction~\cite{vaitukaitis12_petmei}; or off-line user behaviour analysis for game play~\cite{newn2018looks}.
For~\textit{attentive user interfaces}, usually it is sufficient to detect the attention of the user~\cite{alt2016attention} with binary eye contact detection~\cite{smith2013gaze,dickie2004eye,zhang2017everyday,mueller18_etra}.

\paragraph{\textbf{Usability}}
We rate applications with high usability as it can work with calibration-free fashion and can be used with multi-user simultaneously.
Since the \textit{explicit eye input} usually assumes a single target user, it is relatively straightforward to include personal calibration process.
Such a personal calibration is also required from the high accuracy requirement discussed above.
While the~\textit{attentive user interfaces} also requires specific object calibration for each different camera-object relationship~\cite{smith2013gaze}, the underlying use case scenario demands pervasive multi-user gaze estimation without personal calibration.
For~\textit{gaze-based user modelling} and \textit{passive eye monitoring}, relative eye movement could be sufficient considering that there have been already applications implemented in a calibration-free fashion~\cite{zhang13_chi}.
They can be also naturally extended to multi-user scenarios.

\paragraph{\textbf{Robustness}}
Performance consistency between indoor and outdoor environments becomes more and more important with the popularisation of personal mobile devices.
It has impacts on \textit{passive eye monitoring} since usually long-term recording through daily life is necessary~\cite{steil2015discovery}.
\textit{Explicit eye input} and ~\textit{gaze-based user modelling} also require such consistency as users could run these applications anywhere with their mobile devices.
In contrast,~\textit{attentive user interfaces} could be conducted within a stable scene if the target object is stationary.
In addition, there are large numbers of people wearing glasses nowadays.
The glasses could cause problems for gaze-based interaction since thick frames, distortion, and reflection could potentially impair the quality of gaze estimation methods.
For all types of gaze-based applications, they must be robust to the user with or without glasses.

\subsection{Extension of application scenarios}

From our experimental results, we can see that the current appearance-based gaze estimation can achieve reasonable accuracy.
\autoref{fig:results_distances} shows that the appearance-based gaze estimation method can achieve around two to three degrees accuracy, which can provide good enough estimates for some applications.
\autoref{fig:results_samples} and~\autoref{fig:results_indoor_outdoor} show that appearance-based gaze estimation is comparable to Tobii EyeX in terms of its requirements for calibration samples and robustness to indoor and outdoor environments.
Appearance-based gaze estimation could thus be used in applications requiring \textit{explicit eye input}, such as object selection~\cite{d2016gazed,andrist2017looking} or gaze pointing ~\cite{otsuki2017thirdeye,zhang2017smartphone}.
Appearance-based methods are already suitable for applications only requiring measurement of relative changes in gaze direction over time, such as~\textit{gaze-based user modelling},~\textit{passive eye monitoring} or detection of gaze patterns, such as smooth pursuit eye movements~\cite{esteves2015orbits}.
Also the latest~\textit{attentive user interfaces} could use appearance-based gaze estimation methods, such as for eye contact detection~\cite{smith2013gaze,zhang2017everyday,mueller18_etra} or attention forecasting~\cite{steil18_mobilehci}.
This suggests webcams could replace commercial eye trackers for some applications, and even enable new application scenarios such as online software-based services.

From~\autoref{fig:results_distances}, we can see the advantage of appearance-based gaze estimation on large operation at distances between users and camera; it also maintains consistent gaze estimation accuracy across the different distances.
This method enables gaze-based applications for different devices such as a cellphone with a short distance and a large TV with a long distance.
The current gaze-based applications for these short and long distances are limited to rough gaze direction with previous model-based gaze estimation methods~\cite{zhang2017smartphone,zhang2015eye}.
Appearance-based gaze estimation can achieve gaze estimation error around two to three degrees for these devices, which allows researchers to use the estimated gaze points for fine-level interaction, such as object selection or attention measurement.

Another major flaw of current commercial eye trackers is they usually can only work with a single user due to the limited camera angle of view.
This is not an issue for gaze estimation with a webcam, which can output multiple-person gaze information.
Therefore, it enables new applications where multiple persons are involved, and we can achieve their gaze information with a single webcam.
This has been implemented in one previous work that performs the eye contact detection with webcams and there could be more than one user in the input image~\cite{mueller18_etra}.

Above all, gaze estimation with a single webcam instead of an additional commercial eye tracker enables new forms of applications.
It allows researchers to develop gaze-based applications with common devices, such as cellphones, tablets, laptops and TVs.
Participants can stick with their own personal devices and run the gaze-based software to perform the interaction without any additional hardware requirement.
This is the key advantage of using a single webcam for gaze estimation instead of the current commercial eye trackers.
%!TEX root = 00_main.tex

\section{The OpenGaze software toolkit}

As shown before, appearance-based gaze estimation has significant potential to facilitate gaze-based interactive applications on the millions of camera-equipped devices already used worldwide today, such as mobile phones or laptops.
However, most existing methods -- if code is available for these at all -- were published with research-oriented implementations and there is no easy-to-use software toolkit available that is specifically geared to HCI purposes.
It is also challenging for designers to integrate existing computer vision and machine learning pipelines into end-user applications.

We therefore extended the MPIIFaceGaze method into a complete open source toolkit for gaze-based applications.
The goal of our~\toolkitname software toolkit is to provide an easy way for HCI researchers and designers to use appearance-based gaze estimation techniques, and enable a wider range of eye tracking applications using off-the-shelf cameras.
We designed~\toolkitname with four main objectives in mind: 1) to implement state-of-the-art appearance-based gaze estimation for HCI researchers, 2) to make the functionality easy to use and work out-of-the-box for rapid development, 3) to be extensible to include more functions, and 4) to be flexible as developers can replace any components with their own methods.

\begin{figure*}[t]
\centering
  \includegraphics[width=0.9\textwidth]{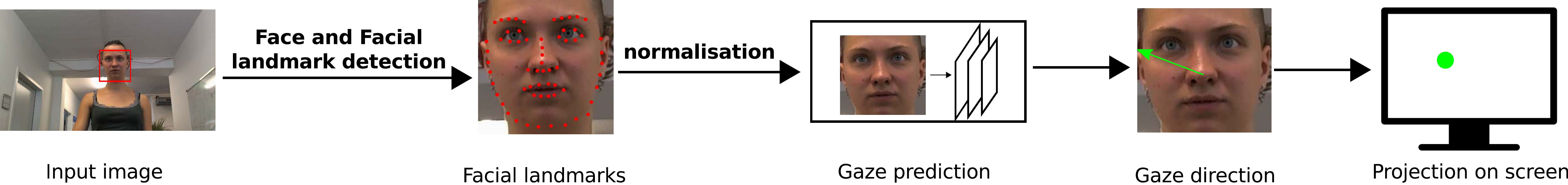}
  \caption{Taking an image as input, our~\toolkitname toolkit first detects the faces and facial landmarks (a) and then crops the face image using data normalisation~\cite{zhang2018revisiting}. The appearance-based gaze estimation model predicts the gaze direction in the camera coordinate system from the normalised face image. The direction is finally converted to the screen coordinate system.}
  \label{fig:opengaze_pipeline}
\end{figure*}

The overall pipeline of~\toolkitname is shown in~\autoref{fig:opengaze_pipeline}.
Unlike the dominate commercial eye trackers, the input of~\toolkitname can be single RGB images, such as the video stream from the camera, recorded videos, a directory of images, or a single image.
Given an input frame/image, our OpenGaze first detects faces and facial landmarks, which are used to estimate 3D head pose and data normalization.
The data normalization procedure essentially crops the face image with a normalised camera to cancel out some of appearance variations caused by head pose.
The cropped face image then will be input to the appearance-based gaze estimation method.
The output of the gaze estimation model is the gaze direction in the camera coordinate system, which can be further projected to the screen coordinate system.
\toolkitname has user-friendly APIs for developers to perform their desired functions with minimal effort on coding, and also provides easy-to-install packages including pre-compiled libraries to facilitate use of gaze estimation in interactive applications.

\subsection{Face and facial landmark detection}

Given an input image, the first step is to detect the face as well as facial landmarks of the user.
\toolkitname integrates OpenFace 2.0~\cite{baltrusaitis2018openface} for facial landmark detection that, in turn, relies on the widely used dlib computer vision library~\cite{dlib09} to detect the faces in the input image.
OpenFace also assigns unique IDs to each detected faces via temporal tracking.
The detected facial landmarks (4 eye corners and 2 mouth corners) are mainly used to estimate 3D head pose including head rotation and translation, which is achieved by fitting a pre-defined generic 3D face model to the detected facial landmarks by estimating the initial solution using the EPnP algorithm~\cite{lepetit2009epnp}.
The estimated 3D head pose is used in data normalisation and the 3D face centre (centroid of the six facial landmarks) is taken as the origin of gaze directions.

\subsection{Data normalisation}
As input for appearance-based gaze estimation, the system then crops and resizes the face image according to the facial landmarks.
However, since appearance-based 3D gaze estimation is a geometric task, inappropriate cropping and resizing can significantly affect the estimation accuracy. 
Data normalization was proposed to efficiently train appearance-based gaze estimators, and it cancels out the geometric variability by warping input images to a normalised space~\cite{zhang2018revisiting}.
\toolkitname implements the data normalization scheme as pre-processing to appearance-based gaze estimation.
Specifically,~\toolkitname first crops the face image after rotating the camera so that the x-axis of the camera coordinate system is perpendicular to the y-axis of the head coordinate system.
Then it scales the image so that the normalised camera is located at a fixed distance away from the face centre.
In this way, the input image has only 2 degrees of freedom in head pose for all kinds of cameras with different intrinsic parameters.

\subsection{Gaze estimation}
\toolkitname implements an appearance-based gaze estimation method that reports state-of-the-art accuracy~\cite{zhang17_cvprw}.
In this method, the whole face image is fed into the convolutional neural network to output 3D gaze directions.
\toolkitname uses the same neural network architecture as in~\cite{zhang17_cvprw}, which is based on the AlexNet architecture~\cite{krizhevsky2012imagenet}.
The toolkit comes with the model used in our experiments, which was pre-trained on two commonly-used gaze datasets with full-face images, MPIIFaceGaze dataset~\cite{zhang17_cvprw} and EYEDIAP dataset~\cite{mora2014eyediap}.
Therefore, while the toolkit is flexible enough to replace the network with user-trained ones, there is basically no need to train the network from scratch, and developers can directly use the pre-trained model.
It is important to note that~\toolkitname is fully extensible, e.g.\ it allows developers to add new network architectures and train models on other datasets.

\subsection{Projection on screen}
The gaze direction is estimated by the appearance-based gaze estimator in the normalised space, and~\toolkitname projects it back to the original camera coordinate system.
\toolkitname also provides APIs to project the 3D gaze direction from the camera coordinate system to the 2D screen coordinate system, and vice versa.
In order to project gaze direction to the 2D screen coordinate system,~\toolkitname requires the camera intrinsic parameters and camera-screen relationship, i.e., rotation and translation between the camera and screen coordinate system.
The camera intrinsic parameters can be obtained by camera calibration function in OpenCV~\cite{opencv_library} by moving a calibration pattern in front of the camera.
The camera-screen relationship can be calculated with a mirror-based camera-screen calibration method~\cite{rodrigues2010camera}.
It requires showing a camera calibration pattern on the screen, and then move a planar mirror in front of the camera to let the camera capture several calibration samples with the full view of the calibration pattern.

\subsection{Personal calibration}
\label{sec:calibration}
Cross-person gaze estimation is the ultimate goal of data-driven appearance-based gaze estimation, and as described above,~\toolkitname comes with a pre-trained generic gaze estimator which works across users, environments, and cameras without any personal calibration~\cite{zhang2017mpiigaze}.
However, if the application allows for additional calibration data collection, the gaze estimation accuracy can be significantly improved by personal calibration.
To make the estimated gaze usable for interactive applications,~\toolkitname further provides a personal calibration scheme to make corrections to raw gaze estimates from the appearance-based gaze estimation model.
To collect the ground-truth calibration samples,~\toolkitname provides a GUI to collect the personal calibration data from users.
During the personal calibration,~\toolkitname shows the shrinking circle on the screen and the user has to fixate on the circle until it become a dot, while confirming that he/she is looking at the dot by mouse-clicking within a half second.
Meanwhile,~\toolkitname also captures the face image from the webcam associated with the dot position on the screen.
These samples are used to find a third-order polynomial mapping function between the estimated and ground-truth 2D gaze locations in the screen coordinate system.

\subsection{Implementation and speed}
With extensibility in mind, we implemented each of the above components as separate classes written in C++ with interfaces to communicate between each other.
It is feasible for developers to replace and include different components if desired.
As far as we tested,~\toolkitname achieved 13 fps at running time with a desktop machine which has a 3.50GHz CPU and a GeForce GTX TITAN Black GPU with 6GB memory with stream from a webcam.
Note the most time-consuming process is the face and facial landmark detection as they can only reach 17 fps.
By replacing these components, which is possible with in our toolkit, it is expected that a much faster speed.
In addition, the gaze estimation network can be also replaced with more compact versions to achieve higher speed and memory efficiency.
%!TEX root = 00_main.tex

\section{Conclusion}

In this work, we compared the appearance-based method with a model-based method and a commercial eye tracker, and we showed that it achieves better performance than the model-based method, and a larger operational range than the commercial eye tracker.
Following the result, we further discussed design implications for the most important gaze-based applications. 
We present the first software toolkit~\toolkitname which provides an easy-to-use webcam-based gaze estimation method for interaction designers.
The goal of the toolkit is to be applied to a diverse range of interactive systems, and we evaluated the performance of state-of-the-art appearance-based gaze estimation across different affordances.
We believe our~\toolkitname enables new forms of application scenarios for both HCI designers and researchers.

\section{Acknowledgments}
This work was supported by the European Research Council (ERC; grant agreement 801708) as well as by
%the Cluster of Excellence on Multimodal Computing and Interaction at Saarland University, Germany,
a JST CREST research grant (JPMJCR14E1), Japan.

%This project has received funding from the European Research Council (ERC) under the European Union's Horizon 2020 research and innovation programme under grant agreement No .

\bibliographystyle{ACM-Reference-Format}
\bibliography{references}


\begin{thebibliography}{79}

%%% the Web address.

\ifx \showCODEN    \undefined \def \showCODEN     #1{\unskip}     \fi
\ifx \showDOI      \undefined \def \showDOI       #1{#1}\fi
\ifx \showISBNx    \undefined \def \showISBNx     #1{\unskip}     \fi
\ifx \showISBNxiii \undefined \def \showISBNxiii  #1{\unskip}     \fi
\ifx \showISSN     \undefined \def \showISSN      #1{\unskip}     \fi
\ifx \showLCCN     \undefined \def \showLCCN      #1{\unskip}     \fi
\ifx \shownote     \undefined \def \shownote      #1{#1}          \fi
\ifx \showarticletitle \undefined \def \showarticletitle #1{#1}   \fi
\ifx \showURL      \undefined \def \showURL       {\relax}        \fi
% invisible to TeX
\providecommand\bibfield[2]{#2}
\providecommand\bibinfo[2]{#2}
\providecommand\natexlab[1]{#1}
\providecommand\showeprint[2][]{arXiv:#2}

\bibitem[\protect\citeauthoryear{Alt, Bulling, Mecke, and Buschek}{Alt
  et~al\mbox{.}}{2016}]%
        {alt2016attention}
\bibfield{author}{\bibinfo{person}{Florian Alt}, \bibinfo{person}{Andreas
  Bulling}, \bibinfo{person}{Lukas Mecke}, {and} \bibinfo{person}{Daniel
  Buschek}.} \bibinfo{year}{2016}\natexlab{}.
\newblock \showarticletitle{Attention, please! Comparing Features for Measuring
  Audience Attention Towards Pervasive Displays}. In
  \bibinfo{booktitle}{\emph{Proc. ACM SIGCHI Conference on Designing
  Interactive Systems (DIS)}}. \bibinfo{pages}{823--828}.
\newblock
\urldef\tempurl%
\url{https://doi.org/10.1145/2901790.2901897}
\showDOI{\tempurl}


\bibitem[\protect\citeauthoryear{Andrist, Gleicher, and Mutlu}{Andrist
  et~al\mbox{.}}{2017}]%
        {andrist2017looking}
\bibfield{author}{\bibinfo{person}{Sean Andrist}, \bibinfo{person}{Michael
  Gleicher}, {and} \bibinfo{person}{Bilge Mutlu}.}
  \bibinfo{year}{2017}\natexlab{}.
\newblock \showarticletitle{Looking Coordinated: Bidirectional Gaze Mechanisms
  for Collaborative Interaction with Virtual Characters}. In
  \bibinfo{booktitle}{\emph{Proceedings of the 2017 CHI Conference on Human
  Factors in Computing Systems}}. ACM, \bibinfo{pages}{2571--2582}.
\newblock


\bibitem[\protect\citeauthoryear{Baltrusaitis, Zadeh, Lim, and
  Morency}{Baltrusaitis et~al\mbox{.}}{2018}]%
        {baltrusaitis2018openface}
\bibfield{author}{\bibinfo{person}{Tadas Baltrusaitis}, \bibinfo{person}{Amir
  Zadeh}, \bibinfo{person}{Yao~Chong Lim}, {and}
  \bibinfo{person}{Louis-Philippe Morency}.} \bibinfo{year}{2018}\natexlab{}.
\newblock \showarticletitle{OpenFace 2.0: Facial Behavior Analysis Toolkit}. In
  \bibinfo{booktitle}{\emph{Automatic Face \& Gesture Recognition (FG 2018),
  2018 13th IEEE International Conference on}}. IEEE, \bibinfo{pages}{59--66}.
\newblock


\bibitem[\protect\citeauthoryear{Bradski}{Bradski}{2000}]%
        {opencv_library}
\bibfield{author}{\bibinfo{person}{G. Bradski}.}
  \bibinfo{year}{2000}\natexlab{}.
\newblock \showarticletitle{{The OpenCV Library}}.
\newblock \bibinfo{journal}{\emph{Dr. Dobb's Journal of Software Tools}}
  (\bibinfo{year}{2000}).
\newblock


\bibitem[\protect\citeauthoryear{Bulling}{Bulling}{2016}]%
        {bulling2016pervasive}
\bibfield{author}{\bibinfo{person}{Andreas Bulling}.}
  \bibinfo{year}{2016}\natexlab{}.
\newblock \showarticletitle{Pervasive Attentive User Interfaces}.
\newblock \bibinfo{journal}{\emph{IEEE Computer}} \bibinfo{volume}{49},
  \bibinfo{number}{1} (\bibinfo{year}{2016}), \bibinfo{pages}{94--98}.
\newblock
\urldef\tempurl%
\url{https://doi.org/10.1109/MC.2016.32}
\showDOI{\tempurl}


\bibitem[\protect\citeauthoryear{Bulling, Weichel, and Gellersen}{Bulling
  et~al\mbox{.}}{2013}]%
        {bulling13_chi}
\bibfield{author}{\bibinfo{person}{Andreas Bulling}, \bibinfo{person}{Christian
  Weichel}, {and} \bibinfo{person}{Hans Gellersen}.}
  \bibinfo{year}{2013}\natexlab{}.
\newblock \showarticletitle{EyeContext: Recognition of High-level Contextual
  Cues from Human Visual Behaviour}. In \bibinfo{booktitle}{\emph{Proc. ACM
  SIGCHI Conference on Human Factors in Computing Systems (CHI)}}.
  \bibinfo{pages}{305--308}.
\newblock
\urldef\tempurl%
\url{https://doi.org/10.1145/2470654.2470697}
\showDOI{\tempurl}


\bibitem[\protect\citeauthoryear{Bulling and Zander}{Bulling and
  Zander}{2014}]%
        {bulling14_pcm}
\bibfield{author}{\bibinfo{person}{Andreas Bulling} {and}
  \bibinfo{person}{Thorsten~O. Zander}.} \bibinfo{year}{2014}\natexlab{}.
\newblock \showarticletitle{Cognition-Aware Computing}.
\newblock \bibinfo{journal}{\emph{IEEE Pervasive Computing}}
  \bibinfo{volume}{13}, \bibinfo{number}{3} (\bibinfo{year}{2014}),
  \bibinfo{pages}{80--83}.
\newblock
\urldef\tempurl%
\url{https://doi.org/10.1109/mprv.2014.42}
\showDOI{\tempurl}


\bibitem[\protect\citeauthoryear{Cantoni, Galdi, Nappi, Porta, and
  Riccio}{Cantoni et~al\mbox{.}}{2015}]%
        {cantoni2015gant}
\bibfield{author}{\bibinfo{person}{Virginio Cantoni}, \bibinfo{person}{Chiara
  Galdi}, \bibinfo{person}{Michele Nappi}, \bibinfo{person}{Marco Porta}, {and}
  \bibinfo{person}{Daniel Riccio}.} \bibinfo{year}{2015}\natexlab{}.
\newblock \showarticletitle{GANT: Gaze analysis technique for human
  identification}.
\newblock \bibinfo{journal}{\emph{Pattern Recognition}} \bibinfo{volume}{48},
  \bibinfo{number}{4} (\bibinfo{year}{2015}), \bibinfo{pages}{1027--1038}.
\newblock
\urldef\tempurl%
\url{https://doi.org/10.1016/j.patcog.2014.02.017}
\showDOI{\tempurl}


\bibitem[\protect\citeauthoryear{Chen and Ji}{Chen and Ji}{2008}]%
        {chen20083d}
\bibfield{author}{\bibinfo{person}{Jixu Chen} {and} \bibinfo{person}{Qiang
  Ji}.} \bibinfo{year}{2008}\natexlab{}.
\newblock \showarticletitle{3D gaze estimation with a single camera without IR
  illumination}. In \bibinfo{booktitle}{\emph{Pattern Recognition, 2008. ICPR
  2008. 19th International Conference on}}. IEEE, \bibinfo{pages}{1--4}.
\newblock


\bibitem[\protect\citeauthoryear{D'Angelo and Gergle}{D'Angelo and
  Gergle}{2016}]%
        {d2016gazed}
\bibfield{author}{\bibinfo{person}{Sarah D'Angelo} {and}
  \bibinfo{person}{Darren Gergle}.} \bibinfo{year}{2016}\natexlab{}.
\newblock \showarticletitle{Gazed and confused: Understanding and designing
  shared gaze for remote collaboration}. In
  \bibinfo{booktitle}{\emph{Proceedings of the 2016 CHI Conference on Human
  Factors in Computing Systems}}. ACM, \bibinfo{pages}{2492--2496}.
\newblock


\bibitem[\protect\citeauthoryear{D'Angelo and Gergle}{D'Angelo and
  Gergle}{2018}]%
        {d2018eye}
\bibfield{author}{\bibinfo{person}{Sarah D'Angelo} {and}
  \bibinfo{person}{Darren Gergle}.} \bibinfo{year}{2018}\natexlab{}.
\newblock \showarticletitle{An Eye For Design: Gaze Visualizations for Remote
  Collaborative Work}. In \bibinfo{booktitle}{\emph{Proceedings of the 2018 CHI
  Conference on Human Factors in Computing Systems}}. ACM,
  \bibinfo{pages}{349}.
\newblock


\bibitem[\protect\citeauthoryear{Dickie, Vertegaal, Shell, Sohn, Cheng, and
  Aoudeh}{Dickie et~al\mbox{.}}{2004}]%
        {dickie2004eye}
\bibfield{author}{\bibinfo{person}{Connor Dickie}, \bibinfo{person}{Roel
  Vertegaal}, \bibinfo{person}{Jeffrey~S Shell}, \bibinfo{person}{Changuk
  Sohn}, \bibinfo{person}{Daniel Cheng}, {and} \bibinfo{person}{Omar Aoudeh}.}
  \bibinfo{year}{2004}\natexlab{}.
\newblock \showarticletitle{Eye contact sensing glasses for attention-sensitive
  wearable video blogging}. In \bibinfo{booktitle}{\emph{CHI'04 extended
  abstracts on Human factors in computing systems}}. ACM,
  \bibinfo{pages}{769--770}.
\newblock


\bibitem[\protect\citeauthoryear{Esteves, Velloso, Bulling, and
  Gellersen}{Esteves et~al\mbox{.}}{2015}]%
        {esteves2015orbits}
\bibfield{author}{\bibinfo{person}{Augusto Esteves}, \bibinfo{person}{Eduardo
  Velloso}, \bibinfo{person}{Andreas Bulling}, {and} \bibinfo{person}{Hans
  Gellersen}.} \bibinfo{year}{2015}\natexlab{}.
\newblock \showarticletitle{Orbits: Enabling Gaze Interaction in Smart Watches
  using Moving Targets}. In \bibinfo{booktitle}{\emph{Proc. ACM Symposium on
  User Interface Software and Technology (UIST)}}. \bibinfo{pages}{457--466}.
\newblock
\urldef\tempurl%
\url{https://doi.org/10.1145/2807442.2807499}
\showDOI{\tempurl}


\bibitem[\protect\citeauthoryear{Faber, Bixler, and D'Mello}{Faber
  et~al\mbox{.}}{2017}]%
        {faber2017automated}
\bibfield{author}{\bibinfo{person}{Myrthe Faber}, \bibinfo{person}{Robert
  Bixler}, {and} \bibinfo{person}{Sidney~K D'Mello}.}
  \bibinfo{year}{2017}\natexlab{}.
\newblock \showarticletitle{An automated behavioral measure of mind wandering
  during computerized reading}.
\newblock \bibinfo{journal}{\emph{Behavior Research Methods}}
  (\bibinfo{year}{2017}), \bibinfo{pages}{1--17}.
\newblock
\urldef\tempurl%
\url{https://doi.org/10.3758/s13428-017-0857-y.}
\showDOI{\tempurl}


\bibitem[\protect\citeauthoryear{Hansen and Ji}{Hansen and Ji}{2010}]%
        {hansen2010eye}
\bibfield{author}{\bibinfo{person}{Dan~Witzner Hansen} {and}
  \bibinfo{person}{Qiang Ji}.} \bibinfo{year}{2010}\natexlab{}.
\newblock \showarticletitle{In the eye of the beholder: A survey of models for
  eyes and gaze}.
\newblock \bibinfo{journal}{\emph{IEEE Transactions on Pattern Analysis and
  Machine Intelligence}} \bibinfo{volume}{32}, \bibinfo{number}{3}
  (\bibinfo{year}{2010}), \bibinfo{pages}{478--500}.
\newblock


\bibitem[\protect\citeauthoryear{Higuch, Yonetani, and Sato}{Higuch
  et~al\mbox{.}}{2016}]%
        {higuch2016can}
\bibfield{author}{\bibinfo{person}{Keita Higuch}, \bibinfo{person}{Ryo
  Yonetani}, {and} \bibinfo{person}{Yoichi Sato}.}
  \bibinfo{year}{2016}\natexlab{}.
\newblock \showarticletitle{Can Eye Help You?: Effects of Visualizing Eye
  Fixations on Remote Collaboration Scenarios for Physical Tasks}. In
  \bibinfo{booktitle}{\emph{Proceedings of the 2016 CHI Conference on Human
  Factors in Computing Systems}}. ACM, \bibinfo{pages}{5180--5190}.
\newblock


\bibitem[\protect\citeauthoryear{Holzman, Proctor, Levy, Yasillo, Meltzer, and
  Hurt}{Holzman et~al\mbox{.}}{1974}]%
        {holzman1974eye}
\bibfield{author}{\bibinfo{person}{Philip~S Holzman},
  \bibinfo{person}{Leonard~R Proctor}, \bibinfo{person}{Deborah~L Levy},
  \bibinfo{person}{Nicholas~J Yasillo}, \bibinfo{person}{Herbert~Y Meltzer},
  {and} \bibinfo{person}{Stephen~W Hurt}.} \bibinfo{year}{1974}\natexlab{}.
\newblock \showarticletitle{Eye-tracking dysfunctions in schizophrenic patients
  and their relatives}.
\newblock \bibinfo{journal}{\emph{Archives of general psychiatry}}
  \bibinfo{volume}{31}, \bibinfo{number}{2} (\bibinfo{year}{1974}),
  \bibinfo{pages}{143--151}.
\newblock


\bibitem[\protect\citeauthoryear{Hoppe, Loetscher, Morey, and Bulling}{Hoppe
  et~al\mbox{.}}{2018}]%
        {hoppe2018eye}
\bibfield{author}{\bibinfo{person}{Sabrina Hoppe}, \bibinfo{person}{Tobias
  Loetscher}, \bibinfo{person}{Stephanie~A Morey}, {and}
  \bibinfo{person}{Andreas Bulling}.} \bibinfo{year}{2018}\natexlab{}.
\newblock \showarticletitle{Eye movements during everyday behavior predict
  personality traits}.
\newblock \bibinfo{journal}{\emph{Frontiers in Human Neuroscience}}
  \bibinfo{volume}{12} (\bibinfo{year}{2018}), \bibinfo{pages}{105}.
\newblock
\urldef\tempurl%
\url{https://doi.org/10.3389/fnhum.2018.00105}
\showDOI{\tempurl}


\bibitem[\protect\citeauthoryear{Huang, Kwok, Ngai, Chan, and Leong}{Huang
  et~al\mbox{.}}{2016a}]%
        {huang2016building}
\bibfield{author}{\bibinfo{person}{Michael~Xuelin Huang},
  \bibinfo{person}{Tiffany~CK Kwok}, \bibinfo{person}{Grace Ngai},
  \bibinfo{person}{Stephen~CF Chan}, {and} \bibinfo{person}{Hong~Va Leong}.}
  \bibinfo{year}{2016}\natexlab{a}.
\newblock \showarticletitle{Building a personalized, auto-calibrating eye
  tracker from user interactions}. In \bibinfo{booktitle}{\emph{Proceedings of
  the 2016 CHI Conference on Human Factors in Computing Systems}}. ACM,
  \bibinfo{pages}{5169--5179}.
\newblock


\bibitem[\protect\citeauthoryear{Huang, Li, Ngai, and Leong}{Huang
  et~al\mbox{.}}{2016b}]%
        {huang2016stressclick}
\bibfield{author}{\bibinfo{person}{Michael~Xuelin Huang},
  \bibinfo{person}{Jiajia Li}, \bibinfo{person}{Grace Ngai}, {and}
  \bibinfo{person}{Hong~Va Leong}.} \bibinfo{year}{2016}\natexlab{b}.
\newblock \showarticletitle{StressClick: Sensing Stress from Gaze-Click
  Patterns}. In \bibinfo{booktitle}{\emph{Proceedings of the 2016 ACM on
  Multimedia Conference}}. ACM, \bibinfo{pages}{1395--1404}.
\newblock


\bibitem[\protect\citeauthoryear{Huang, Li, Ngai, and Leong}{Huang
  et~al\mbox{.}}{2017}]%
        {huang2017screenglint}
\bibfield{author}{\bibinfo{person}{Michael~Xuelin Huang},
  \bibinfo{person}{Jiajia Li}, \bibinfo{person}{Grace Ngai}, {and}
  \bibinfo{person}{Hong~Va Leong}.} \bibinfo{year}{2017}\natexlab{}.
\newblock \showarticletitle{Screenglint: Practical, in-situ gaze estimation on
  smartphones}. In \bibinfo{booktitle}{\emph{Proceedings of the 2017 CHI
  Conference on Human Factors in Computing Systems}}. ACM,
  \bibinfo{pages}{2546--2557}.
\newblock


\bibitem[\protect\citeauthoryear{Huang, Veeraraghavan, and Sabharwal}{Huang
  et~al\mbox{.}}{2015}]%
        {huang2015tabletgaze}
\bibfield{author}{\bibinfo{person}{Qiong Huang}, \bibinfo{person}{Ashok
  Veeraraghavan}, {and} \bibinfo{person}{Ashutosh Sabharwal}.}
  \bibinfo{year}{2015}\natexlab{}.
\newblock \showarticletitle{TabletGaze: unconstrained appearance-based gaze
  estimation in mobile tablets}.
\newblock \bibinfo{journal}{\emph{arXiv preprint arXiv:1508.01244}}
  (\bibinfo{year}{2015}).
\newblock


\bibitem[\protect\citeauthoryear{Hutton, Nagel, and Loewenson}{Hutton
  et~al\mbox{.}}{1984}]%
        {hutton1984eye}
\bibfield{author}{\bibinfo{person}{J~Thomas Hutton}, \bibinfo{person}{JA
  Nagel}, {and} \bibinfo{person}{Ruth~B Loewenson}.}
  \bibinfo{year}{1984}\natexlab{}.
\newblock \showarticletitle{Eye tracking dysfunction in Alzheimer-type
  dementia}.
\newblock \bibinfo{journal}{\emph{Neurology}} \bibinfo{volume}{34},
  \bibinfo{number}{1} (\bibinfo{year}{1984}), \bibinfo{pages}{99--99}.
\newblock


\bibitem[\protect\citeauthoryear{Ishikawa, Baker, Matthews, and
  Kanade}{Ishikawa et~al\mbox{.}}{2004}]%
        {ishikawa2004passive}
\bibfield{author}{\bibinfo{person}{Takahiro Ishikawa}, \bibinfo{person}{Simon
  Baker}, \bibinfo{person}{Iain Matthews}, {and} \bibinfo{person}{Takeo
  Kanade}.} \bibinfo{year}{2004}\natexlab{}.
\newblock \showarticletitle{Passive driver gaze tracking with active appearance
  models}. In \bibinfo{booktitle}{\emph{Proceedings of the 11th world congress
  on intelligent transportation systems}}, Vol.~\bibinfo{volume}{3}.
  \bibinfo{pages}{41--43}.
\newblock


\bibitem[\protect\citeauthoryear{Khamis, Alt, and Bulling}{Khamis
  et~al\mbox{.}}{2018}]%
        {khamis18_mobilehci}
\bibfield{author}{\bibinfo{person}{Mohamed Khamis}, \bibinfo{person}{Florian
  Alt}, {and} \bibinfo{person}{Andreas Bulling}.}
  \bibinfo{year}{2018}\natexlab{}.
\newblock \showarticletitle{The Past, Present, and Future of Gaze-enabled
  Handheld Mobile Devices: Survey and Lessons Learned}. In
  \bibinfo{booktitle}{\emph{Proc. International Conference on Human-Computer
  Interaction with Mobile Devices and Services (MobileHCI)}}.
  \bibinfo{pages}{38:1--38:17}.
\newblock
\urldef\tempurl%
\url{https://doi.org/10.1145/3229434.3229452}
\showDOI{\tempurl}
\newblock
\shownote{best paper honourable mention award.}


\bibitem[\protect\citeauthoryear{Khamis, Alt, Hassib, von Zezschwitz,
  Hasholzner, and Bulling}{Khamis et~al\mbox{.}}{2016}]%
        {khamis16_chi}
\bibfield{author}{\bibinfo{person}{Mohamed Khamis}, \bibinfo{person}{Florian
  Alt}, \bibinfo{person}{Mariam Hassib}, \bibinfo{person}{Emanuel von
  Zezschwitz}, \bibinfo{person}{Regina Hasholzner}, {and}
  \bibinfo{person}{Andreas Bulling}.} \bibinfo{year}{2016}\natexlab{}.
\newblock \showarticletitle{GazeTouchPass: Multimodal Authentication Using Gaze
  and Touch on Mobile Devices}. In \bibinfo{booktitle}{\emph{Ext. Abstr. ACM
  SIGCHI Conference on Human Factors in Computing Systems (CHI)}}.
  \bibinfo{pages}{2156--2164}.
\newblock
\urldef\tempurl%
\url{https://doi.org/10.1145/2851581.2892314}
\showDOI{\tempurl}


\bibitem[\protect\citeauthoryear{King}{King}{2009}]%
        {dlib09}
\bibfield{author}{\bibinfo{person}{Davis~E. King}.}
  \bibinfo{year}{2009}\natexlab{}.
\newblock \showarticletitle{Dlib-ml: A Machine Learning Toolkit}.
\newblock \bibinfo{journal}{\emph{Journal of Machine Learning Research}}
  \bibinfo{volume}{10} (\bibinfo{year}{2009}), \bibinfo{pages}{1755--1758}.
\newblock


\bibitem[\protect\citeauthoryear{Kosch, Hassib, Wozniak, Buschek, and
  Alt}{Kosch et~al\mbox{.}}{2018}]%
        {kosch2018your}
\bibfield{author}{\bibinfo{person}{Thomas Kosch}, \bibinfo{person}{Mariam
  Hassib}, \bibinfo{person}{Pawel~W Wozniak}, \bibinfo{person}{Daniel Buschek},
  {and} \bibinfo{person}{Florian Alt}.} \bibinfo{year}{2018}\natexlab{}.
\newblock \showarticletitle{Your Eyes Tell: Leveraging Smooth Pursuit for
  Assessing Cognitive Workload}. In \bibinfo{booktitle}{\emph{Proceedings of
  the 2018 CHI Conference on Human Factors in Computing Systems}}. ACM,
  \bibinfo{pages}{436}.
\newblock


\bibitem[\protect\citeauthoryear{Krizhevsky, Sutskever, and Hinton}{Krizhevsky
  et~al\mbox{.}}{2012}]%
        {krizhevsky2012imagenet}
\bibfield{author}{\bibinfo{person}{Alex Krizhevsky}, \bibinfo{person}{Ilya
  Sutskever}, {and} \bibinfo{person}{Geoffrey~E Hinton}.}
  \bibinfo{year}{2012}\natexlab{}.
\newblock \showarticletitle{Imagenet classification with deep convolutional
  neural networks}. In \bibinfo{booktitle}{\emph{Advances in neural information
  processing systems}}. \bibinfo{pages}{1097--1105}.
\newblock


\bibitem[\protect\citeauthoryear{Kuechenmeister, Linton, Mueller, and
  White}{Kuechenmeister et~al\mbox{.}}{1977}]%
        {kuechenmeister1977eye}
\bibfield{author}{\bibinfo{person}{Craig~A Kuechenmeister},
  \bibinfo{person}{Patrick~H Linton}, \bibinfo{person}{Thelma~V Mueller}, {and}
  \bibinfo{person}{Hilton~B White}.} \bibinfo{year}{1977}\natexlab{}.
\newblock \showarticletitle{Eye tracking in relation to age, sex, and illness}.
\newblock \bibinfo{journal}{\emph{Archives of General Psychiatry}}
  \bibinfo{volume}{34}, \bibinfo{number}{5} (\bibinfo{year}{1977}),
  \bibinfo{pages}{578--579}.
\newblock


\bibitem[\protect\citeauthoryear{Kurauchi, Feng, Joshi, Morimoto, and
  Betke}{Kurauchi et~al\mbox{.}}{2016}]%
        {kurauchi2016eyeswipe}
\bibfield{author}{\bibinfo{person}{Andrew Kurauchi}, \bibinfo{person}{Wenxin
  Feng}, \bibinfo{person}{Ajjen Joshi}, \bibinfo{person}{Carlos Morimoto},
  {and} \bibinfo{person}{Margrit Betke}.} \bibinfo{year}{2016}\natexlab{}.
\newblock \showarticletitle{EyeSwipe: Dwell-free text entry using gaze paths}.
  In \bibinfo{booktitle}{\emph{Proceedings of the 2016 CHI Conference on Human
  Factors in Computing Systems}}. ACM, \bibinfo{pages}{1952--1956}.
\newblock


\bibitem[\protect\citeauthoryear{Lagun, Hsieh, Webster, and Navalpakkam}{Lagun
  et~al\mbox{.}}{2014}]%
        {lagun2014towards}
\bibfield{author}{\bibinfo{person}{Dmitry Lagun}, \bibinfo{person}{Chih-Hung
  Hsieh}, \bibinfo{person}{Dale Webster}, {and} \bibinfo{person}{Vidhya
  Navalpakkam}.} \bibinfo{year}{2014}\natexlab{}.
\newblock \showarticletitle{Towards better measurement of attention and
  satisfaction in mobile search}. In \bibinfo{booktitle}{\emph{Proceedings of
  the 37th international ACM SIGIR conference on Research \& development in
  information retrieval}}. ACM, \bibinfo{pages}{113--122}.
\newblock


\bibitem[\protect\citeauthoryear{Lepetit, Moreno-Noguer, and Fua}{Lepetit
  et~al\mbox{.}}{2009}]%
        {lepetit2009epnp}
\bibfield{author}{\bibinfo{person}{Vincent Lepetit}, \bibinfo{person}{Francesc
  Moreno-Noguer}, {and} \bibinfo{person}{Pascal Fua}.}
  \bibinfo{year}{2009}\natexlab{}.
\newblock \showarticletitle{Epnp: An accurate o (n) solution to the pnp
  problem}.
\newblock \bibinfo{journal}{\emph{International journal of computer vision}}
  \bibinfo{volume}{81}, \bibinfo{number}{2} (\bibinfo{year}{2009}),
  \bibinfo{pages}{155}.
\newblock


\bibitem[\protect\citeauthoryear{Li, Xu, Lagun, and Navalpakkam}{Li
  et~al\mbox{.}}{2017}]%
        {li2017towards}
\bibfield{author}{\bibinfo{person}{Yixuan Li}, \bibinfo{person}{Pingmei Xu},
  \bibinfo{person}{Dmitry Lagun}, {and} \bibinfo{person}{Vidhya Navalpakkam}.}
  \bibinfo{year}{2017}\natexlab{}.
\newblock \showarticletitle{Towards measuring and inferring user interest from
  gaze}. In \bibinfo{booktitle}{\emph{Proceedings of the 26th International
  Conference on World Wide Web Companion}}. International World Wide Web
  Conferences Steering Committee, \bibinfo{pages}{525--533}.
\newblock


\bibitem[\protect\citeauthoryear{Majaranta, Ahola, and {\v{S}}pakov}{Majaranta
  et~al\mbox{.}}{2009}]%
        {majaranta2009fast}
\bibfield{author}{\bibinfo{person}{P{\"a}ivi Majaranta},
  \bibinfo{person}{Ulla-Kaija Ahola}, {and} \bibinfo{person}{Oleg
  {\v{S}}pakov}.} \bibinfo{year}{2009}\natexlab{}.
\newblock \showarticletitle{Fast gaze typing with an adjustable dwell time}. In
  \bibinfo{booktitle}{\emph{Proceedings of the SIGCHI Conference on Human
  Factors in Computing Systems}}. ACM, \bibinfo{pages}{357--360}.
\newblock


\bibitem[\protect\citeauthoryear{Majaranta and Bulling}{Majaranta and
  Bulling}{2014}]%
        {majaranta2014eye}
\bibfield{author}{\bibinfo{person}{P{\"a}ivi Majaranta} {and}
  \bibinfo{person}{Andreas Bulling}.} \bibinfo{year}{2014}\natexlab{}.
\newblock \showarticletitle{Eye tracking and eye-based human--computer
  interaction}.
\newblock In \bibinfo{booktitle}{\emph{Advances in physiological computing}}.
  \bibinfo{publisher}{Springer}, \bibinfo{pages}{39--65}.
\newblock


\bibitem[\protect\citeauthoryear{Mardanbegi, Hansen, and Pederson}{Mardanbegi
  et~al\mbox{.}}{2012}]%
        {mardanbegi2012eye}
\bibfield{author}{\bibinfo{person}{Diako Mardanbegi},
  \bibinfo{person}{Dan~Witzner Hansen}, {and} \bibinfo{person}{Thomas
  Pederson}.} \bibinfo{year}{2012}\natexlab{}.
\newblock \showarticletitle{Eye-based head gestures}. In
  \bibinfo{booktitle}{\emph{Proceedings of the symposium on eye tracking
  research and applications}}. ACM, \bibinfo{pages}{139--146}.
\newblock


\bibitem[\protect\citeauthoryear{Matthews, Middleton, Gilmartin, and
  Bullimore}{Matthews et~al\mbox{.}}{1991}]%
        {matthews1991pupillary}
\bibfield{author}{\bibinfo{person}{G Matthews}, \bibinfo{person}{W Middleton},
  \bibinfo{person}{B Gilmartin}, {and} \bibinfo{person}{MA Bullimore}.}
  \bibinfo{year}{1991}\natexlab{}.
\newblock \showarticletitle{Pupillary diameter and cognitive load.}
\newblock \bibinfo{journal}{\emph{Journal of Psychophysiology}}
  (\bibinfo{year}{1991}).
\newblock


\bibitem[\protect\citeauthoryear{Mora, Monay, and Odobez}{Mora
  et~al\mbox{.}}{2014}]%
        {mora2014eyediap}
\bibfield{author}{\bibinfo{person}{Kenneth Alberto~Funes Mora},
  \bibinfo{person}{Florent Monay}, {and} \bibinfo{person}{Jean-Marc Odobez}.}
  \bibinfo{year}{2014}\natexlab{}.
\newblock \showarticletitle{Eyediap: A database for the development and
  evaluation of gaze estimation algorithms from rgb and rgb-d cameras}. In
  \bibinfo{booktitle}{\emph{Proceedings of the Symposium on Eye Tracking
  Research and Applications}}. ACM, \bibinfo{pages}{255--258}.
\newblock


\bibitem[\protect\citeauthoryear{Mott, Williams, Wobbrock, and Morris}{Mott
  et~al\mbox{.}}{2017}]%
        {mott2017improving}
\bibfield{author}{\bibinfo{person}{Martez~E Mott}, \bibinfo{person}{Shane
  Williams}, \bibinfo{person}{Jacob~O Wobbrock}, {and}
  \bibinfo{person}{Meredith~Ringel Morris}.} \bibinfo{year}{2017}\natexlab{}.
\newblock \showarticletitle{Improving dwell-based gaze typing with dynamic,
  cascading dwell times}. In \bibinfo{booktitle}{\emph{Proceedings of the 2017
  CHI Conference on Human Factors in Computing Systems}}. ACM,
  \bibinfo{pages}{2558--2570}.
\newblock


\bibitem[\protect\citeauthoryear{M{\"u}ller, Huang, and Bulling}{M{\"u}ller
  et~al\mbox{.}}{2018a}]%
        {muller2018detecting}
\bibfield{author}{\bibinfo{person}{Philipp M{\"u}ller},
  \bibinfo{person}{Michael~Xuelin Huang}, {and} \bibinfo{person}{Andreas
  Bulling}.} \bibinfo{year}{2018}\natexlab{a}.
\newblock \showarticletitle{Detecting Low Rapport During Natural Interactions
  in Small Groups from Non-Verbal Behaviour}. In \bibinfo{booktitle}{\emph{23rd
  International Conference on Intelligent User Interfaces}}. ACM,
  \bibinfo{pages}{153--164}.
\newblock


\bibitem[\protect\citeauthoryear{M{\"u}ller, Huang, Zhang, and
  Bulling}{M{\"u}ller et~al\mbox{.}}{2018b}]%
        {mueller18_etra}
\bibfield{author}{\bibinfo{person}{Philipp M{\"u}ller},
  \bibinfo{person}{Michael~Xuelin Huang}, \bibinfo{person}{Xucong Zhang}, {and}
  \bibinfo{person}{Andreas Bulling}.} \bibinfo{year}{2018}\natexlab{b}.
\newblock \showarticletitle{Robust Eye Contact Detection in Natural
  Multi-Person Interactions Using Gaze and Speaking Behaviour}. In
  \bibinfo{booktitle}{\emph{Proc. International Symposium on Eye Tracking
  Research and Applications (ETRA)}}. \bibinfo{pages}{31:1--31:10}.
\newblock
\urldef\tempurl%
\url{https://doi.org/10.1145/3204493.3204549}
\showDOI{\tempurl}


\bibitem[\protect\citeauthoryear{Newn, Allison, Velloso, and Vetere}{Newn
  et~al\mbox{.}}{2018}]%
        {newn2018looks}
\bibfield{author}{\bibinfo{person}{Joshua Newn}, \bibinfo{person}{Fraser
  Allison}, \bibinfo{person}{Eduardo Velloso}, {and} \bibinfo{person}{Frank
  Vetere}.} \bibinfo{year}{2018}\natexlab{}.
\newblock \showarticletitle{Looks can be deceiving: Using gaze visualisation to
  predict and mislead opponents in strategic gameplay}. In
  \bibinfo{booktitle}{\emph{Proceedings of the 2018 CHI Conference on Human
  Factors in Computing Systems}}. ACM, \bibinfo{pages}{261}.
\newblock


\bibitem[\protect\citeauthoryear{Nguyen and Liu}{Nguyen and Liu}{2016}]%
        {nguyen2016gaze}
\bibfield{author}{\bibinfo{person}{Cuong Nguyen} {and} \bibinfo{person}{Feng
  Liu}.} \bibinfo{year}{2016}\natexlab{}.
\newblock \showarticletitle{Gaze-based Notetaking for Learning from Lecture
  Videos}. In \bibinfo{booktitle}{\emph{Proceedings of the 2016 CHI Conference
  on Human Factors in Computing Systems}}. ACM, \bibinfo{pages}{2093--2097}.
\newblock


\bibitem[\protect\citeauthoryear{Otsuki, Kawano, Maruyama, Kuzuoka, and
  Suzuki}{Otsuki et~al\mbox{.}}{2017}]%
        {otsuki2017thirdeye}
\bibfield{author}{\bibinfo{person}{Mai Otsuki}, \bibinfo{person}{Taiki Kawano},
  \bibinfo{person}{Keita Maruyama}, \bibinfo{person}{Hideaki Kuzuoka}, {and}
  \bibinfo{person}{Yusuke Suzuki}.} \bibinfo{year}{2017}\natexlab{}.
\newblock \showarticletitle{ThirdEye: Simple Add-on Display to Represent Remote
  Participant's Gaze Direction in Video Communication}. In
  \bibinfo{booktitle}{\emph{Proceedings of the 2017 CHI Conference on Human
  Factors in Computing Systems}}. ACM, \bibinfo{pages}{5307--5312}.
\newblock


\bibitem[\protect\citeauthoryear{Palinko, Kun, Shyrokov, and Heeman}{Palinko
  et~al\mbox{.}}{2010}]%
        {palinko2010estimating}
\bibfield{author}{\bibinfo{person}{Oskar Palinko}, \bibinfo{person}{Andrew~L
  Kun}, \bibinfo{person}{Alexander Shyrokov}, {and} \bibinfo{person}{Peter
  Heeman}.} \bibinfo{year}{2010}\natexlab{}.
\newblock \showarticletitle{Estimating cognitive load using remote eye tracking
  in a driving simulator}. In \bibinfo{booktitle}{\emph{Proceedings of the 2010
  symposium on eye-tracking research \& applications}}. ACM,
  \bibinfo{pages}{141--144}.
\newblock


\bibitem[\protect\citeauthoryear{Park, Zhang, Bulling, and Hilliges}{Park
  et~al\mbox{.}}{2018}]%
        {park18_etra}
\bibfield{author}{\bibinfo{person}{Seonwook Park}, \bibinfo{person}{Xucong
  Zhang}, \bibinfo{person}{Andreas Bulling}, {and} \bibinfo{person}{Otmar
  Hilliges}.} \bibinfo{year}{2018}\natexlab{}.
\newblock \showarticletitle{Learning to Find Eye Region Landmarks for Remote
  Gaze Estimation in Unconstrained Settings}. In
  \bibinfo{booktitle}{\emph{Proc. International Symposium on Eye Tracking
  Research and Applications (ETRA)}}. \bibinfo{pages}{21:1--21:10}.
\newblock
\urldef\tempurl%
\url{https://doi.org/10.1145/3204493.3204545}
\showDOI{\tempurl}


\bibitem[\protect\citeauthoryear{Piumsomboon, Lee, Lindeman, and
  Billinghurst}{Piumsomboon et~al\mbox{.}}{2017}]%
        {piumsomboon2017exploring}
\bibfield{author}{\bibinfo{person}{Thammathip Piumsomboon},
  \bibinfo{person}{Gun Lee}, \bibinfo{person}{Robert~W Lindeman}, {and}
  \bibinfo{person}{Mark Billinghurst}.} \bibinfo{year}{2017}\natexlab{}.
\newblock \showarticletitle{Exploring natural eye-gaze-based interaction for
  immersive virtual reality}. In \bibinfo{booktitle}{\emph{3D User Interfaces
  (3DUI), 2017 IEEE Symposium on}}. IEEE, \bibinfo{pages}{36--39}.
\newblock


\bibitem[\protect\citeauthoryear{Rodrigues, Barreto, and Nunes}{Rodrigues
  et~al\mbox{.}}{2010}]%
        {rodrigues2010camera}
\bibfield{author}{\bibinfo{person}{Rui Rodrigues}, \bibinfo{person}{Joao~P.
  Barreto}, {and} \bibinfo{person}{Urbano Nunes}.}
  \bibinfo{year}{2010}\natexlab{}.
\newblock \showarticletitle{Camera pose estimation using images of planar
  mirror reflections}. In \bibinfo{booktitle}{\emph{Proceedings of the 11th
  European Conference on Computer Vision}}. \bibinfo{pages}{382--395}.
\newblock


\bibitem[\protect\citeauthoryear{Sammaknejad, Pouretemad, Eslahchi, Salahirad,
  and Alinejad}{Sammaknejad et~al\mbox{.}}{2017}]%
        {sammaknejad2017gender}
\bibfield{author}{\bibinfo{person}{Negar Sammaknejad},
  \bibinfo{person}{Hamidreza Pouretemad}, \bibinfo{person}{Changiz Eslahchi},
  \bibinfo{person}{Alireza Salahirad}, {and} \bibinfo{person}{Ashkan
  Alinejad}.} \bibinfo{year}{2017}\natexlab{}.
\newblock \showarticletitle{Gender classification based on eye movements: A
  processing effect during passive face viewing}.
\newblock \bibinfo{journal}{\emph{Advances in cognitive psychology}}
  \bibinfo{volume}{13}, \bibinfo{number}{3} (\bibinfo{year}{2017}),
  \bibinfo{pages}{232}.
\newblock


\bibitem[\protect\citeauthoryear{Sattar, M{\"{u}}ller, Fritz, and
  Bulling}{Sattar et~al\mbox{.}}{2015}]%
        {sattar15_cvpr}
\bibfield{author}{\bibinfo{person}{Hosnieh Sattar}, \bibinfo{person}{Sabine
  M{\"{u}}ller}, \bibinfo{person}{Mario Fritz}, {and} \bibinfo{person}{Andreas
  Bulling}.} \bibinfo{year}{2015}\natexlab{}.
\newblock \showarticletitle{Prediction of Search Targets From Fixations in
  Open-world Settings}. In \bibinfo{booktitle}{\emph{Proc. IEEE Conference on
  Computer Vision and Pattern Recognition (CVPR)}}. \bibinfo{pages}{981--990}.
\newblock
\urldef\tempurl%
\url{https://doi.org/10.1109/CVPR.2015.7298700}
\showDOI{\tempurl}


\bibitem[\protect\citeauthoryear{Schenk, Dreiser, Rigoll, and Dorr}{Schenk
  et~al\mbox{.}}{2017}]%
        {schenk2017gazeeverywhere}
\bibfield{author}{\bibinfo{person}{Simon Schenk}, \bibinfo{person}{Marc
  Dreiser}, \bibinfo{person}{Gerhard Rigoll}, {and} \bibinfo{person}{Michael
  Dorr}.} \bibinfo{year}{2017}\natexlab{}.
\newblock \showarticletitle{GazeEverywhere: Enabling Gaze-only User Interaction
  on an Unmodified Desktop PC in Everyday Scenarios}. In
  \bibinfo{booktitle}{\emph{Proceedings of the 2017 CHI Conference on Human
  Factors in Computing Systems}}. ACM, \bibinfo{pages}{3034--3044}.
\newblock


\bibitem[\protect\citeauthoryear{Shrivastava, Pfister, Tuzel, Susskind, Wang,
  and Webb}{Shrivastava et~al\mbox{.}}{2017}]%
        {Shrivastava2016Learning}
\bibfield{author}{\bibinfo{person}{Ashish Shrivastava}, \bibinfo{person}{Tomas
  Pfister}, \bibinfo{person}{Oncel Tuzel}, \bibinfo{person}{Josh Susskind},
  \bibinfo{person}{Wenda Wang}, {and} \bibinfo{person}{Russ Webb}.}
  \bibinfo{year}{2017}\natexlab{}.
\newblock \showarticletitle{Learning from Simulated and Unsupervised Images
  through Adversarial Training}. In \bibinfo{booktitle}{\emph{Computer Vision
  and Pattern Recognition (CVPR), 2017 IEEE Conference on}}.
\newblock


\bibitem[\protect\citeauthoryear{Smith, Yin, Feiner, and Nayar}{Smith
  et~al\mbox{.}}{2013}]%
        {smith2013gaze}
\bibfield{author}{\bibinfo{person}{Brian~A Smith}, \bibinfo{person}{Qi Yin},
  \bibinfo{person}{Steven~K Feiner}, {and} \bibinfo{person}{Shree~K Nayar}.}
  \bibinfo{year}{2013}\natexlab{}.
\newblock \showarticletitle{Gaze locking: passive eye contact detection for
  human-object interaction}. In \bibinfo{booktitle}{\emph{Proceedings of the
  26th annual ACM symposium on User interface software and technology}}. ACM,
  \bibinfo{pages}{271--280}.
\newblock


\bibitem[\protect\citeauthoryear{Steil and Bulling}{Steil and Bulling}{2015}]%
        {steil2015discovery}
\bibfield{author}{\bibinfo{person}{Julian Steil} {and} \bibinfo{person}{Andreas
  Bulling}.} \bibinfo{year}{2015}\natexlab{}.
\newblock \showarticletitle{Discovery of everyday human activities from
  long-term visual behaviour using topic models}. In
  \bibinfo{booktitle}{\emph{Proceedings of the 2015 ACM International Joint
  Conference on Pervasive and Ubiquitous Computing}}. ACM,
  \bibinfo{pages}{75--85}.
\newblock
\urldef\tempurl%
\url{https://doi.org/10.1145/2750858.2807520}
\showDOI{\tempurl}


\bibitem[\protect\citeauthoryear{Steil, Müller, Sugano, and Bulling}{Steil
  et~al\mbox{.}}{2018}]%
        {steil18_mobilehci}
\bibfield{author}{\bibinfo{person}{Julian Steil}, \bibinfo{person}{Philipp
  Müller}, \bibinfo{person}{Yusuke Sugano}, {and} \bibinfo{person}{Andreas
  Bulling}.} \bibinfo{year}{2018}\natexlab{}.
\newblock \showarticletitle{Forecasting User Attention During Everyday Mobile
  Interactions Using Device-Integrated and Wearable Sensors}. In
  \bibinfo{booktitle}{\emph{Proc. International Conference on Human-Computer
  Interaction with Mobile Devices and Services (MobileHCI)}} (2018-04-16).
  \bibinfo{pages}{1:1--1:13}.
\newblock
\urldef\tempurl%
\url{https://doi.org/10.1145/3229434.3229439}
\showDOI{\tempurl}


\bibitem[\protect\citeauthoryear{Sugano, Zhang, and Bulling}{Sugano
  et~al\mbox{.}}{2016}]%
        {sugano2016aggregaze}
\bibfield{author}{\bibinfo{person}{Yusuke Sugano}, \bibinfo{person}{Xucong
  Zhang}, {and} \bibinfo{person}{Andreas Bulling}.}
  \bibinfo{year}{2016}\natexlab{}.
\newblock \showarticletitle{Aggregaze: Collective estimation of audience
  attention on public displays}. In \bibinfo{booktitle}{\emph{Proceedings of
  the 29th Annual Symposium on User Interface Software and Technology}}. ACM,
  \bibinfo{pages}{821--831}.
\newblock


\bibitem[\protect\citeauthoryear{Tan, Kriegman, and Ahuja}{Tan
  et~al\mbox{.}}{2002}]%
        {tan2002appearance}
\bibfield{author}{\bibinfo{person}{Kar-Han Tan}, \bibinfo{person}{David~J
  Kriegman}, {and} \bibinfo{person}{Narendra Ahuja}.}
  \bibinfo{year}{2002}\natexlab{}.
\newblock \showarticletitle{Appearance-based eye gaze estimation}. In
  \bibinfo{booktitle}{\emph{Applications of Computer Vision, 2002.(WACV 2002).
  Proceedings. Sixth IEEE Workshop on}}. IEEE, \bibinfo{pages}{191--195}.
\newblock


\bibitem[\protect\citeauthoryear{Tessendorf, Bulling, Roggen, Stiefmeier,
  Feilner, Derleth, and Tr{\"{o}}ster}{Tessendorf et~al\mbox{.}}{2011}]%
        {tessendorf11_pervasive}
\bibfield{author}{\bibinfo{person}{Bernd Tessendorf}, \bibinfo{person}{Andreas
  Bulling}, \bibinfo{person}{Daniel Roggen}, \bibinfo{person}{Thomas
  Stiefmeier}, \bibinfo{person}{Manuela Feilner}, \bibinfo{person}{Peter
  Derleth}, {and} \bibinfo{person}{Gerhard Tr{\"{o}}ster}.}
  \bibinfo{year}{2011}\natexlab{}.
\newblock \showarticletitle{Recognition of Hearing Needs From Body and Eye
  Movements to Improve Hearing Instruments}. In \bibinfo{booktitle}{\emph{Proc.
  International Conference on Pervasive Computing (Pervasive)}}.
  \bibinfo{pages}{314--331}.
\newblock
\urldef\tempurl%
\url{https://doi.org/10.1007/978-3-642-21726-5_20}
\showDOI{\tempurl}


\bibitem[\protect\citeauthoryear{Vaitukaitis and Bulling}{Vaitukaitis and
  Bulling}{2012}]%
        {vaitukaitis12_petmei}
\bibfield{author}{\bibinfo{person}{Vytautas Vaitukaitis} {and}
  \bibinfo{person}{Andreas Bulling}.} \bibinfo{year}{2012}\natexlab{}.
\newblock \showarticletitle{Eye Gesture Recognition on Portable Devices}. In
  \bibinfo{booktitle}{\emph{Proc. International Workshop on Pervasive Eye
  Tracking and Mobile Gaze-Based Interaction (PETMEI)}}.
  \bibinfo{pages}{711--714}.
\newblock
\urldef\tempurl%
\url{https://doi.org/10.1145/2370216.2370370}
\showDOI{\tempurl}


\bibitem[\protect\citeauthoryear{Valenti, Sebe, and Gevers}{Valenti
  et~al\mbox{.}}{2012}]%
        {valenti2012combining}
\bibfield{author}{\bibinfo{person}{Roberto Valenti}, \bibinfo{person}{Nicu
  Sebe}, {and} \bibinfo{person}{Theo Gevers}.} \bibinfo{year}{2012}\natexlab{}.
\newblock \showarticletitle{Combining head pose and eye location information
  for gaze estimation}.
\newblock \bibinfo{journal}{\emph{IEEE Transactions on Image Processing}}
  \bibinfo{volume}{21}, \bibinfo{number}{2} (\bibinfo{year}{2012}),
  \bibinfo{pages}{802--815}.
\newblock


\bibitem[\protect\citeauthoryear{Vertegaal et~al\mbox{.}}{Vertegaal
  et~al\mbox{.}}{2003}]%
        {vertegaal2003attentive}
\bibfield{author}{\bibinfo{person}{Roel Vertegaal} {et~al\mbox{.}}}
  \bibinfo{year}{2003}\natexlab{}.
\newblock \showarticletitle{Attentive user interfaces}.
\newblock \bibinfo{journal}{\emph{Commun. ACM}} \bibinfo{volume}{46},
  \bibinfo{number}{3} (\bibinfo{year}{2003}), \bibinfo{pages}{30--33}.
\newblock


\bibitem[\protect\citeauthoryear{Vidal, Bulling, and Gellersen}{Vidal
  et~al\mbox{.}}{2013}]%
        {vidal13_ubicomp}
\bibfield{author}{\bibinfo{person}{M{\'{e}}lodie Vidal},
  \bibinfo{person}{Andreas Bulling}, {and} \bibinfo{person}{Hans Gellersen}.}
  \bibinfo{year}{2013}\natexlab{}.
\newblock \showarticletitle{Pursuits: Spontaneous Interaction with Displays
  based on Smooth Pursuit Eye Movement and Moving Targets}. In
  \bibinfo{booktitle}{\emph{Proc. ACM International Joint Conference on
  Pervasive and Ubiquitous Computing (UbiComp)}}. \bibinfo{pages}{439--448}.
\newblock
\urldef\tempurl%
\url{https://doi.org/10.1145/2468356.2479632}
\showDOI{\tempurl}


\bibitem[\protect\citeauthoryear{Wood, Baltrusaitis, Zhang, Sugano, Robinson,
  and Bulling}{Wood et~al\mbox{.}}{2015}]%
        {wood2015rendering}
\bibfield{author}{\bibinfo{person}{Erroll Wood}, \bibinfo{person}{Tadas
  Baltrusaitis}, \bibinfo{person}{Xucong Zhang}, \bibinfo{person}{Yusuke
  Sugano}, \bibinfo{person}{Peter Robinson}, {and} \bibinfo{person}{Andreas
  Bulling}.} \bibinfo{year}{2015}\natexlab{}.
\newblock \showarticletitle{Rendering of eyes for eye-shape registration and
  gaze estimation}. In \bibinfo{booktitle}{\emph{Proceedings of the IEEE
  International Conference on Computer Vision}}. \bibinfo{pages}{3756--3764}.
\newblock


\bibitem[\protect\citeauthoryear{Wood and Bulling}{Wood and Bulling}{2014}]%
        {wood2014eyetab}
\bibfield{author}{\bibinfo{person}{Erroll Wood} {and} \bibinfo{person}{Andreas
  Bulling}.} \bibinfo{year}{2014}\natexlab{}.
\newblock \showarticletitle{Eyetab: Model-based gaze estimation on unmodified
  tablet computers}. In \bibinfo{booktitle}{\emph{Proceedings of the Symposium
  on Eye Tracking Research and Applications}}. ACM, \bibinfo{pages}{207--210}.
\newblock


\bibitem[\protect\citeauthoryear{Xu, Sugano, and Bulling}{Xu
  et~al\mbox{.}}{2016}]%
        {xu2016spatio}
\bibfield{author}{\bibinfo{person}{Pingmei Xu}, \bibinfo{person}{Yusuke
  Sugano}, {and} \bibinfo{person}{Andreas Bulling}.}
  \bibinfo{year}{2016}\natexlab{}.
\newblock \showarticletitle{Spatio-temporal modeling and prediction of visual
  attention in graphical user interfaces}. In
  \bibinfo{booktitle}{\emph{Proceedings of the 2016 CHI Conference on Human
  Factors in Computing Systems}}. ACM, \bibinfo{pages}{3299--3310}.
\newblock


\bibitem[\protect\citeauthoryear{Yamazoe, Utsumi, Yonezawa, and Abe}{Yamazoe
  et~al\mbox{.}}{2008}]%
        {yamazoe2008remote}
\bibfield{author}{\bibinfo{person}{Hirotake Yamazoe}, \bibinfo{person}{Akira
  Utsumi}, \bibinfo{person}{Tomoko Yonezawa}, {and} \bibinfo{person}{Shinji
  Abe}.} \bibinfo{year}{2008}\natexlab{}.
\newblock \showarticletitle{Remote gaze estimation with a single camera based
  on facial-feature tracking without special calibration actions}. In
  \bibinfo{booktitle}{\emph{Proceedings of the 2008 symposium on Eye tracking
  research \& applications}}. ACM, \bibinfo{pages}{245--250}.
\newblock


\bibitem[\protect\citeauthoryear{Zhang, Huang, Sugano, and Bulling}{Zhang
  et~al\mbox{.}}{2018a}]%
        {zhang2018training}
\bibfield{author}{\bibinfo{person}{Xucong Zhang},
  \bibinfo{person}{Michael~Xuelin Huang}, \bibinfo{person}{Yusuke Sugano},
  {and} \bibinfo{person}{Andreas Bulling}.} \bibinfo{year}{2018}\natexlab{a}.
\newblock \showarticletitle{Training Person-Specific Gaze Estimators from
  Interactions with Multiple Devices}. In \bibinfo{booktitle}{\emph{Proc. ACM
  SIGCHI Conference on Human Factors in Computing Systems (CHI)}}.
  \bibinfo{pages}{624:1--624:12}.
\newblock
\urldef\tempurl%
\url{https://doi.org/10.1145/3173574.3174198}
\showDOI{\tempurl}


\bibitem[\protect\citeauthoryear{Zhang, Kulkarni, and Morris}{Zhang
  et~al\mbox{.}}{2017a}]%
        {zhang2017smartphone}
\bibfield{author}{\bibinfo{person}{Xiaoyi Zhang}, \bibinfo{person}{Harish
  Kulkarni}, {and} \bibinfo{person}{Meredith~Ringel Morris}.}
  \bibinfo{year}{2017}\natexlab{a}.
\newblock \showarticletitle{Smartphone-Based Gaze Gesture Communication for
  People with Motor Disabilities}. In \bibinfo{booktitle}{\emph{Proceedings of
  the 2017 CHI Conference on Human Factors in Computing Systems}}. ACM,
  \bibinfo{pages}{2878--2889}.
\newblock


\bibitem[\protect\citeauthoryear{Zhang, Sugano, and Bulling}{Zhang
  et~al\mbox{.}}{2017c}]%
        {zhang2017everyday}
\bibfield{author}{\bibinfo{person}{Xucong Zhang}, \bibinfo{person}{Yusuke
  Sugano}, {and} \bibinfo{person}{Andreas Bulling}.}
  \bibinfo{year}{2017}\natexlab{c}.
\newblock \showarticletitle{Everyday Eye Contact Detection Using Unsupervised
  Gaze Target Discovery}. In \bibinfo{booktitle}{\emph{Proc. ACM Symposium on
  User Interface Software and Technology (UIST)}}. \bibinfo{pages}{193--203}.
\newblock
\urldef\tempurl%
\url{https://doi.org/10.1145/3126594.3126614}
\showDOI{\tempurl}


\bibitem[\protect\citeauthoryear{Zhang, Sugano, and Bulling}{Zhang
  et~al\mbox{.}}{2018b}]%
        {zhang2018revisiting}
\bibfield{author}{\bibinfo{person}{Xucong Zhang}, \bibinfo{person}{Yusuke
  Sugano}, {and} \bibinfo{person}{Andreas Bulling}.}
  \bibinfo{year}{2018}\natexlab{b}.
\newblock \showarticletitle{Revisiting data normalization for appearance-based
  gaze estimation}. In \bibinfo{booktitle}{\emph{Proceedings of the 2018 ACM
  Symposium on Eye Tracking Research \& Applications}}. ACM,
  \bibinfo{pages}{12}.
\newblock


\bibitem[\protect\citeauthoryear{Zhang, Sugano, Fritz, and Bulling}{Zhang
  et~al\mbox{.}}{2017d}]%
        {zhang17_cvprw}
\bibfield{author}{\bibinfo{person}{Xucong Zhang}, \bibinfo{person}{Yusuke
  Sugano}, \bibinfo{person}{Mario Fritz}, {and} \bibinfo{person}{Andreas
  Bulling}.} \bibinfo{year}{2017}\natexlab{d}.
\newblock \showarticletitle{It's written all over your face: Full-face
  appearance-based gaze estimation}. In \bibinfo{booktitle}{\emph{Computer
  Vision and Pattern Recognition Workshops (CVPRW), 2017 IEEE Conference on}}.
  IEEE, \bibinfo{pages}{2299--2308}.
\newblock


\bibitem[\protect\citeauthoryear{Zhang, Sugano, Fritz, and Bulling}{Zhang
  et~al\mbox{.}}{2018c}]%
        {zhang2017mpiigaze}
\bibfield{author}{\bibinfo{person}{Xucong Zhang}, \bibinfo{person}{Yusuke
  Sugano}, \bibinfo{person}{Mario Fritz}, {and} \bibinfo{person}{Andreas
  Bulling}.} \bibinfo{year}{2018}\natexlab{c}.
\newblock \showarticletitle{MPIIGaze: Real-world dataset and deep
  appearance-based gaze estimation}.
\newblock \bibinfo{journal}{\emph{IEEE Transactions on Pattern Analysis and
  Machine Intelligence}} (\bibinfo{year}{2018}).
\newblock


\bibitem[\protect\citeauthoryear{Zhang, Bulling, and Gellersen}{Zhang
  et~al\mbox{.}}{2013}]%
        {zhang13_chi}
\bibfield{author}{\bibinfo{person}{Yanxia Zhang}, \bibinfo{person}{Andreas
  Bulling}, {and} \bibinfo{person}{Hans Gellersen}.}
  \bibinfo{year}{2013}\natexlab{}.
\newblock \showarticletitle{SideWays: A Gaze Interface for Spontaneous
  Interaction with Situated Displays}. In \bibinfo{booktitle}{\emph{Proc. ACM
  SIGCHI Conference on Human Factors in Computing Systems (CHI)}}.
  \bibinfo{pages}{851--860}.
\newblock
\urldef\tempurl%
\url{https://doi.org/10.1145/2470654.2470775}
\showDOI{\tempurl}


\bibitem[\protect\citeauthoryear{Zhang, Chong, M\"uller, Bulling, and
  Gellersen}{Zhang et~al\mbox{.}}{2015}]%
        {zhang2015eye}
\bibfield{author}{\bibinfo{person}{Yanxia Zhang}, \bibinfo{person}{Ming~Ki
  Chong}, \bibinfo{person}{J\"org M\"uller}, \bibinfo{person}{Andreas Bulling},
  {and} \bibinfo{person}{Hans Gellersen}.} \bibinfo{year}{2015}\natexlab{}.
\newblock \showarticletitle{Eye tracking for public displays in the wild}.
\newblock \bibinfo{journal}{\emph{Springer Personal and Ubiquitous Computing}}
  \bibinfo{volume}{19}, \bibinfo{number}{5} (\bibinfo{year}{2015}),
  \bibinfo{pages}{967--981}.
\newblock
\urldef\tempurl%
\url{https://doi.org/10.1007/s00779-015-0866-8}
\showDOI{\tempurl}


\bibitem[\protect\citeauthoryear{Zhang, M{\"{u}}ller, Chong, Bulling, and
  Gellersen}{Zhang et~al\mbox{.}}{2014}]%
        {zhang14_ubicomp}
\bibfield{author}{\bibinfo{person}{Yanxia Zhang},
  \bibinfo{person}{Hans~J{\"{o}}rg M{\"{u}}ller}, \bibinfo{person}{Ming~Ki
  Chong}, \bibinfo{person}{Andreas Bulling}, {and} \bibinfo{person}{Hans
  Gellersen}.} \bibinfo{year}{2014}\natexlab{}.
\newblock \showarticletitle{GazeHorizon: Enabling Passers-by to Interact with
  Public Displays by Gaze}. In \bibinfo{booktitle}{\emph{Proc. ACM
  International Joint Conference on Pervasive and Ubiquitous Computing
  (UbiComp)}}. \bibinfo{pages}{559--563}.
\newblock
\urldef\tempurl%
\url{https://doi.org/10.1145/2632048.2636071}
\showDOI{\tempurl}


\bibitem[\protect\citeauthoryear{Zhang, Pfeuffer, Chong, Alexander, Bulling,
  and Gellersen}{Zhang et~al\mbox{.}}{2017b}]%
        {zhang2017look}
\bibfield{author}{\bibinfo{person}{Yanxia Zhang}, \bibinfo{person}{Ken
  Pfeuffer}, \bibinfo{person}{Ming~Ki Chong}, \bibinfo{person}{Jason
  Alexander}, \bibinfo{person}{Andreas Bulling}, {and} \bibinfo{person}{Hans
  Gellersen}.} \bibinfo{year}{2017}\natexlab{b}.
\newblock \showarticletitle{Look together: using gaze for assisting co-located
  collaborative search}.
\newblock \bibinfo{journal}{\emph{Personal and Ubiquitous Computing}}
  \bibinfo{volume}{21}, \bibinfo{number}{1} (\bibinfo{year}{2017}),
  \bibinfo{pages}{173--186}.
\newblock


\bibitem[\protect\citeauthoryear{Zhu and Ji}{Zhu and Ji}{2005}]%
        {zhu2005eye}
\bibfield{author}{\bibinfo{person}{Zhiwei Zhu} {and} \bibinfo{person}{Qiang
  Ji}.} \bibinfo{year}{2005}\natexlab{}.
\newblock \showarticletitle{Eye gaze tracking under natural head movements}. In
  \bibinfo{booktitle}{\emph{Computer Vision and Pattern Recognition, 2005. CVPR
  2005. IEEE Computer Society Conference on}}, Vol.~\bibinfo{volume}{1}. IEEE,
  \bibinfo{pages}{918--923}.
\newblock


\bibitem[\protect\citeauthoryear{Zhu, Ji, and Bennett}{Zhu
  et~al\mbox{.}}{2006}]%
        {zhu2006nonlinear}
\bibfield{author}{\bibinfo{person}{Zhiwei Zhu}, \bibinfo{person}{Qiang Ji},
  {and} \bibinfo{person}{Kristin~P Bennett}.} \bibinfo{year}{2006}\natexlab{}.
\newblock \showarticletitle{Nonlinear eye gaze mapping function estimation via
  support vector regression}. In \bibinfo{booktitle}{\emph{Pattern Recognition,
  2006. ICPR 2006. 18th International Conference on}},
  Vol.~\bibinfo{volume}{1}. IEEE, \bibinfo{pages}{1132--1135}.
\newblock


\end{thebibliography}

\end{document}